\documentclass{aastex61}
\pdfoutput=1

\usepackage{amsfonts}
\usepackage{amssymb}
\usepackage{epsfig}
\usepackage{courier}
\usepackage{graphicx}
\usepackage{amsmath}

\newcommand{\be}{\begin{equation}}
\newcommand{\ee}{\end{equation}}
\newcommand{\bea}{\begin{eqnarray}}
\newcommand{\eea}{\end{eqnarray}}

\newcommand{\ti}{\times}

\newcommand{\mc}{\mathcal}

\newcommand{\beqa}{\begin{eqnarray}}
\newcommand{\eeqa}{\end{eqnarray}}

\received{March, 2017}

\submitjournal{ApJ}
\shorttitle{ALP constraints from NGC1275}
\shortauthors{Berg et al.}

\begin{document}

\title{Constraints on Axion-Like Particles from\\ X-ray Observations of NGC1275}

\author{Marcus Berg}
\affiliation{Department of Physics, Karlstad University, 651 88 Karlstad, Sweden}
\author{Joseph P.~Conlon}
\affiliation{Rudolf Peierls Centre for Theoretical Physics, 1 Keble Road, Oxford, OX1 3NP, UK}
\author{Francesca Day}
\affiliation{Rudolf Peierls Centre for Theoretical Physics, 1 Keble Road, Oxford, OX1 3NP, UK}
\author{ Nicholas Jennings}
\affiliation{Rudolf Peierls Centre for Theoretical Physics, 1 Keble Road, Oxford, OX1 3NP, UK}
\author{Sven Krippendorf}
\affiliation{Rudolf Peierls Centre for Theoretical Physics, 1 Keble Road, Oxford, OX1 3NP, UK}
\author{Andrew J.~Powell}
\affiliation{Rudolf Peierls Centre for Theoretical Physics, 1 Keble Road, Oxford, OX1 3NP, UK}
\author{Markus Rummel}
\affiliation{Rudolf Peierls Centre for Theoretical Physics, 1 Keble Road, Oxford, OX1 3NP, UK}
%\emailAdd{marcus.berg@kau.se}
%\emailAdd{joseph.conlon@physics.ox.ac.uk}
%\emailAdd{francesca.day@physics.ox.ac.uk}
%\emailAdd{nicholas.jennings@physics.ox.ac.uk}
%\emailAdd{sven.krippendorf@physics.ox.ac.uk}
%\emailAdd{andrew.powell2@physics.ox.ac.uk}
%\emailAdd{markus.rummel@physics.ox.ac.uk}

%\maketitle

\begin{abstract}
Axion-like particles (ALPs) can induce localised oscillatory modulations in the spectra of photon sources passing through
astrophysical magnetic fields. Ultra-deep \emph{Chandra} observations of the Perseus cluster contain
over $5 \ti 10^5$ counts from the AGN of the central cluster galaxy NGC1275, and represent a dataset of extraordinary quality for ALP searches.
We use this dataset to search for X-ray spectral irregularities from the AGN. The absence of irregularities at
the $\mc{O}(30 \%)$ level allows us to place leading constraints on the ALP-photon mixing
parameter $g_{a\gamma\gamma} \lesssim 1.4 - 4.0 \ti 10^{-12} {\rm GeV}^{-1}$ for $m_a \lesssim 10^{-12} {\rm eV}$, depending on assumptions on the
magnetic field realisation along the line of sight.
\end{abstract}
%%\correspondingauthor{Sven Krippendorf}
%%\email{sven.krippendorf@physics.ox.ac.uk}

%\preprint{OUTP-YYYY}
%\newpage
\tableofcontents

\section{Introduction}

The Perseus galaxy cluster (A426) is the brightest X-ray cluster in the sky. It is a cool-core cluster at redshift $z=0.0176$, centered
around the Seyfert galaxy NGC1275 and its Active Galactic Nucleus (AGN).
Due to its proximity and brightness, the Perseus cluster has been a standard target for all X-ray satellites.
The X-ray spectrum and emission specifically from the AGN are described in~\cite{Churazov:2003hr, Yamazaki, Balmaverde2006, Fabian:2015kua}.

Axion-like particles (ALPs) are a well-motivated extension of the Standard Model, and arise generically in string compactifications (for example see
\cite{hep-th/0602233, hep-th/0605206, 1206.0819}). As the potential and interactions of ALPs are protected by shift symmetries,
they can naturally have extremely small masses or even be massless.\footnote{The shift symmetry also implies that, even if massless,
ALPs are not constrained by searches for fifth forces or modifications of general relativity.}
If it exists, an ALP $a$ interacts with the Standard Model via a coupling to electromagnetism,
\be
\label{ALPphoton}
\frac{a}{M} {\bf E} \cdot {\bf B}~,
\ee
where $M \equiv g_{a\gamma\gamma}^{-1}$ parametrises the strength of the interaction
and ${\bf E}$ and ${\bf B}$ are the electric and magnetic fields. A general review of ALPs and their physics can be found in
\cite{RingwaldReview}.
We concern ourselves in this paper with the case $m_a \lesssim 10^{-12} {\rm eV}$, for which
non-observation of gamma ray photons coincident with the
SN1987A neutrino burst constrains $M \gtrsim 2 \cdot 10^{11} {\rm GeV}$~\cite{astro-ph/9605197,astro-ph/9606028,1410.3747}.

If ALPs exist, then the interaction of equation (\ref{ALPphoton})
causes ALPs and photons to interconvert in the presence of a background magnetic field $\langle B \rangle$~\cite{Sikivie:1983ip, Raffelt:1987im}.
Starting with a pure photon spectrum, this $\gamma \leftrightarrow a$ interconversion results in modulations in the spectrum of
arriving photons.

Galaxy clusters are particularly efficient photon-ALP converters~\cite{0902.2320, 1305.3603}, and
for the electron densities and magnetic fields present within galaxy clusters, it is a result that
at X-ray energies
the $\gamma \leftrightarrow a$ conversion probability is both energy-dependent and quasi-sinusoidal~\cite{1304.0989, 1305.3603, 1312.3947, 1509.06748}.
Compared to the source spectrum, the spectrum of arriving
photons then has oscillatory modulations imprinted on it. By searching
for such modulations, we can place constraints on the coupling parameter $M$.

For this purpose, quasars or AGNs that are either behind or embedded in galaxy clusters provide attractive sources.
The original photon spectrum is reasonably well described by an absorbed power-law, and all photons arise
from a single sightline passing through the cluster.
As bright sources, AGNs can also provide the large number of counts necessary for
statistical leverage in searching for oscillatory modulations of the photon spectrum.

This method is indeed already largely described and was used in~\cite{1304.0989} (see also~\cite{1205.6428, BraxWoutersBrun}, and~\cite{1603.06978} for
a recent analysis of NGC1275 in gamma rays using this approach). However,~\cite{1304.0989} only
applied these ideas to the study of the AGN at the centre of the Hydra A galaxy cluster (redshift $z=0.052$), for which 200ks of
\emph{Chandra} observation time exist. But in this case, the combination of the intrinsic AGN brightness and the
redshift of $z=0.052$ results in only
a few thousand counts in total, limiting the ability to produce bounds.

In contrast, the AGN at the centre of NGC1275 is both exceedingly bright and the subject of enormous observational time.
NGC1275 has been observed for 1 Ms by \emph{Chandra} ACIS-S, with observations taken in 2002 and 2004, and also for
a further 0.5 Ms by \emph{Chandra} ACIS-I in 2009. Taken together, these generate over half a million X-ray counts from the central AGN
-- a dataset of extraordinary quality for searching for spectral irregularities. Furthermore, there are also 180 ks of observation time with \emph{XMM-Newton} taken in 2001 and 2006. This dataset is not as rich as the \emph{Chandra} dataset, but allows us to cross check our analysis with a different instrument.
In this paper, we use these datasets to search for, and constrain, axion-like particles.

The paper is organised as follows. Section~\ref{sec:ALPS} provides further details on the physics of ALPs and the attractiveness of bright quasars or AGNs
for searching for ALPs. Section~\ref{sec:observations} describes the \emph{Chandra} and \emph{XMM-Newton} observations that we have used and their properties. Section~\ref{sec:analysis} describes the \emph{Chandra} analysis,
the effects of pile-up and strategies taken to mitigate pile-up. Section \ref{XMMAnalysis_sec} describes the analysis of \emph{XMM-Newton} data. In Section~\ref{sec:bounds} we describe bounds on the ALP-photon coupling and in Section~\ref{sec:conclusions} we conclude.
In Appendix~\ref{sec:signal} we discuss irregularities in the data at the 10\% level and mention possible instrumental, astrophysical or new physics explanations.%%CHANGE

\section{AGNs and ALP physics}
\label{sec:ALPS}

If ALPs exist, they interconvert with photons in a background magnetic field $\langle B \rangle \neq 0$~\cite{Sikivie:1983ip,Sikivie:1985,Raffelt:1987im}.
This occurs because the magnetic field
generates, via equation (\ref{ALPphoton}), a 2-particle interaction between
the ALP $a$ and the photon $\gamma$, resulting in a mixing of the mass eigenstates.
In a mathematically identical fashion to neutrino oscillations, photons and ALPs then
have a finite probability of inter-conversion as they pass through the magnetic field.
This is a quantum-mechanical effect, and results in an evolution of the quantum state
$$
|\psi_{init}\rangle = | \gamma(E) \rangle \longrightarrow |\psi_{final} \rangle = \alpha |\gamma(E) \rangle + \beta |a(E) \rangle~,
$$
where $\vert \alpha \vert^2 + \vert \beta \vert^2 = 1$, and $\gamma(E) \left[ a(E) \right]$ denotes a photon [ALP] with energy $E$.
The conversion probability $\vert \beta \vert^2$ depends on the free electron density (which sets the effective photon mass), the magnetic field (which sets the
strength of the mixing) and the magnetic field coherence length (which determines the region over which mixing applies).

The calculational details of this are standard.
Following the original papers~\cite{Sikivie:1983ip, Sikivie:1985, Raffelt:1987im}, the physics of ALPs passing through
magnetic fields has been
described in many works. An incomplete list of articles studying aspects of photon-ALP interconversion in astrophysical magnetic fields
includes~\cite{Brockway:1996yr, Grifols:1996id, 0901.4085, 0902.2320, 0911.0015,1202.6529,1204.6187,1207.0776, 1302.1208,1304.0989, 1305.3603, 1312.3947, 1410.3747,1411.4172, 1412.4777,Schlederer, 1509.06748}.

In this respect, galaxy clusters at X-ray energies sit at a sweet spot for photon-ALP physics.
This is due to two key results. First, galaxy clusters are \emph{particularly efficient} environments
for photon-ALP interconversion. The electron densities are relatively low.
Clusters have magnetic fields that are not significantly smaller than in galaxies, but in which
the ${\bf B}$-field extends over megaparsec scales, far greater than the tens of kiloparsecs
applicable for galactic magnetic fields. The magnetic field coherence lengths in clusters are also larger than in galaxies, comfortably
reaching tens of kiloparsecs.
For massless ALPs, this feature singles out galaxy clusters as providing the most suitable environment
in the universe for ALP-photon interconversion.\footnote{Although they appear appealing, magnetars and related objects do not provide efficient
environments for ALP-photon conversion~\cite{Raffelt:1987im}.}

For convenience we will restrict to massless ALPs in this paper. As the efficiency of ALP-photon
conversion depends on $\vert m_a^2 - \omega_{pl}^2 \vert^{-2},$ where $\omega_{pl}$
is the plasma frequency in galaxy clusters, in practice all masses $m_a \lesssim 10^{-12} {\rm eV}$ are equivalent for our purposes. For $m_a \sim 10^{-12} {\rm eV}$, resonant conversion can occur, allowing weaker ALP-photon couplings to be probed,
but we will not consider that case explicitly in this paper.

\begin{figure}
\centering
\includegraphics[width=0.49\textwidth]{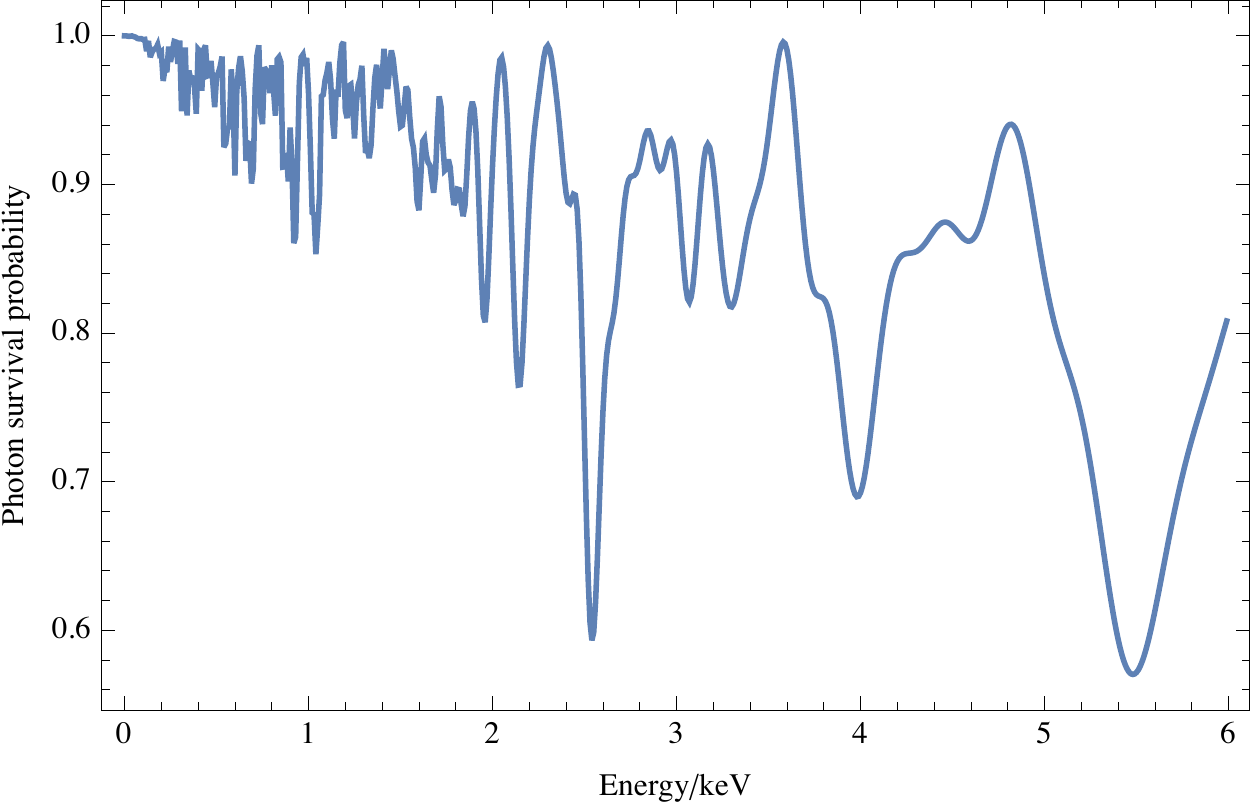} \includegraphics[width=0.49\textwidth]{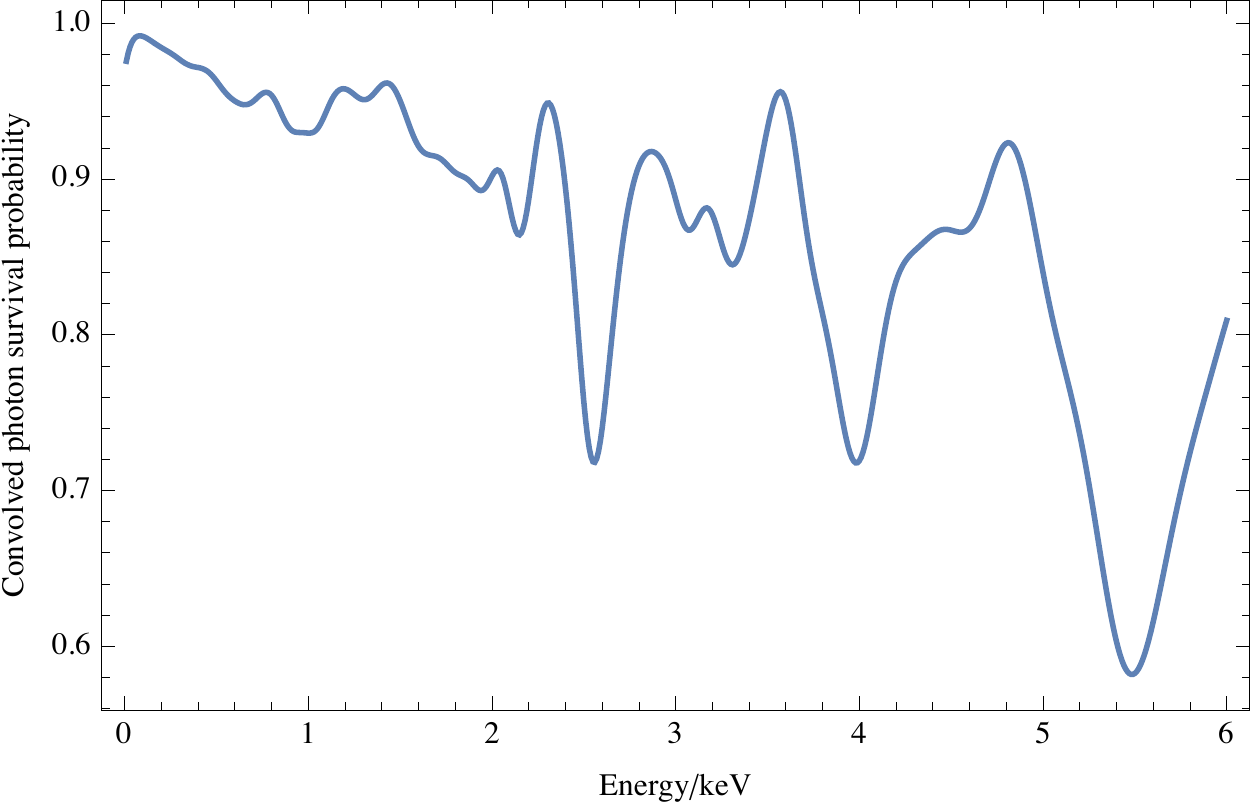}
\caption{Left---The photon survival probability along a line of sight modelled on that from NGC1275 to us,
for a randomly generated magnetic field. A central magnetic field of $B_0 = 25 \mu G$ was assumed, with a radial
scaling of $\langle B(r) \rangle \sim n_e(r)^{0.7}$. There were 200 domains, with lengths drawn randomly between 3.5 and 10 kpc, and the total propagation length being
1200 kpc. The ALP-photon coupling is $g_{a\gamma\gamma} = 1.5 \ti 10^{-12} {\rm GeV}^{-1}$ (roughly a factor of three beyond the current
upper limit $g_{a \gamma \gamma} < 5 \ti 10^{-12} {\rm GeV}^{-1}$ from SN1987A).
One should not take too seriously the particular details of the magnetic field
model used, as there is no observational data to constrain the precise range of magnetic field coherence lengths in the Perseus cluster, but the quasi-sinusoidal structure arises generically.
Right---The photon survival probability for the same magnetic field convolved with a Gaussian with FWHM of 150eV.
 \label{FullALPPhotonConversion}}
\end{figure}

The second key result is that, for the electron densities and magnetic field structures present within
 galaxy clusters, the photon-ALP conversion probability is energy-dependent, with a
quasi-sinusoidal oscillatory structure at X-ray energies. This provides distinctive spectral features to search for.
We illustrate this in Figure~\ref{FullALPPhotonConversion}, where we plot a typical photon survival
probability as a function of energy, along a single line of sight modelled on that from NGC1275 to us.

The precise form of the survival probability is not predictable. It depends on the
actual magnetic field structure along the line of sight, and so differs for each line of sight. Faraday rotation measures can give
statistical information about the strength and extent of magnetic fields, as for the Coma cluster in~\cite{1002.0594}, but
the actual magnetic field along any one line of sight is unknown. However the form shown in Figure~\ref{FullALPPhotonConversion} -- a quasi-sinusoidal structure with a
period that increases with energy -- is generic, and arises for any reasonable choice of central
magnetic field value or range of coherence lengths. The inefficiency of conversion at energies $E \lesssim 0.2 {\rm keV}$ is also generic, implying that
effects of photon-ALP conversion are not visible in the optical (and below) range.

Figure~\ref{FullALPPhotonConversion} (right) also shows the same survival probability convolved with a Gaussian with full width at half maximum (FWHM) of $150{\rm eV},$ representing
the approximate energy resolution of the CCD detectors present on \emph{Chandra} and \emph{XMM-Newton} satellites (the precise figure of $150 {\rm eV}$ is taken
from the in-orbit performance of the ACIS-I detectors on \emph{Chandra}, see table~6.4 of the \emph{Chandra} proposer's
guide\footnote{{\tt http://cxc.harvard.edu/proposer/POG/html/chap6.html}}).
While at lowest energies the oscillations are too rapid to be resolved by CCD detectors, and would require micro-calorimeters
such as those that were present on \emph{Hitomi}\footnote{The energy resolution from \emph{Hitomi} was around $\sim$ 5~eV.}, in general
it is fortuitous that the scales of the oscillations match those
of the X-ray telescopes extensively used to observe galaxy clusters.

If ALPs exist, then
for photons arriving from a single location, this conversion imprints a
particular quasi-sinusoidal modulation on the actual photon spectrum. There is also an overall reduction in luminosity, but this
 can be absorbed into the overall normalisation of the spectrum. For unpolarised light the $\gamma \to a$ conversion probability cannot
exceed fifty per cent, and in the limit of strong coupling saturates at an average value of $\langle P(\gamma \to a) \rangle = 1/3$ (for example, see
\cite{1205.6428}).
It therefore follows that, expressed as a ratio of data to model, the maximal allowed range of ALP-induced modulations is approximately
$\pm 30 \%$.\footnote{We re-emphasise here that photon-ALP conversion involves quantum oscillations between states rather than absorption.
Therefore for passage from $A \to C$ the survival probability $P(A \to C)$ does \emph{not} equal $P(A \to B) \ti P(B \to C)$.}

There are three effects that can wash out these modulations.
The first is the finite energy resolution of the telescope; as shown in
Figure~\ref{FullALPPhotonConversion}, this removes any structure present at the lowest energies.
The second is when emission arises from an extended source, involving many different lines of sight. In this case the
peaks and troughs from different lines of sight
undergo destructive interference, reducing any signal.
The third is insufficient photon statistics, when localised oscillations become indistinguishable from Poisson fluctuations.

Bright point-like sources either behind or embedded in a galaxy cluster are particularly attractive for searching for ALP-induced
modulations. The galaxy cluster provides a good environment for ALP-photon conversion; the bright point source ensures
there are many photons, all passing along the same line of sight.

These factors make quasar or AGN spectra attractive for searching for ALPs.
Emission from an active galactic nucleus (AGN) arises from matter accreting onto the central black hole.
As evidenced by the rapid time variability of AGN luminosities, the physical region sourcing the X-ray AGN
emission is tiny -- of order a few
Schwarschild radii of the central black hole. As cluster magnetic fields are ordered on kiloparsec scales, this implies
that for all practical purposes every photon arising from the AGN has experienced an identical magnetic field structure
during its passage to us.

To first approximation, at X-ray energies an AGN spectrum can be described as an absorbed power-law. The effect of ALPs is then
to imprint a quasi-sinusoidal modulation on this power law, of relative amplitude at most $\mc{O}(30 \%)$
and with a modulation period of order a few hundred eV.
 As the fractional Poisson error on $N$ counts is $\frac{1}{\sqrt{N}}$, and CCD detectors such as those
 on \emph{Chandra} and \emph{XMM-Newton} have intrinsic energy resolutions of around $\mc{O}(100 {\rm eV})$,
 it therefore requires large numbers of counts to be able to distinguish any ALP-induced modulations from
normal statistical fluctuations.

All the above facts make
the AGN of the Seyfert galaxy NGC1275 an excellent candidate for searching for
ALP-photon interconversion. NGC1275 is the central galaxy of the Perseus cluster, which as a cool core cluster
should have a high central magnetic field (estimated as 25 $\mu G$ in~\cite{0602622}) -- implying the sightline from
NGC1275 to us should be efficient at ALP-photon conversion.

The nucleus is intrinsically bright and unobscured, with a spectrum that is well characterised by
a power-law and narrow Fe K$\alpha$ line, absorbed by the galactic $n_H$ column density~\cite{Churazov:2003hr}.
Furthermore, there is enormous \emph{Chandra} observation time on NGC1275, encompassing 1.5 Ms in total. This
results in over half a million photon counts originating from the central AGN, although for the on-axis observations quite a number are contaminated by pile-up.
This is a level two orders of magnitude larger than the study in~\cite{1304.0989} involving Hydra A.

\section{The Observations}
\label{sec:observations}

\subsection{Chandra}

The deep Chandra observations involving NGC1275 can be divided into three main groups. The first involves 200ks of ACIS-S observations
taken in 2002 together with 800ks of ACIS-S observations taken in 2004.\footnote{Note for particle theorists: the ACIS instrument has two main modes, ACIS-S
and ACIS-I. One ACIS-S chip leads to an 8 arcminute by 8 arcminute field of view, while ACIS-I will result in a
16 arcminute by 16 arcminute field of view.} In these observations NGC1275 is close to the aimpoint.
The second group involves 300ks of ACIS-I observations carried out in 2009, where NGC1275 is approximately midway between the edge of the
chips and the aimpoint. The third group involves 200ks
of ACIS-I observations also taken in 2009, in which NGC1275 is close to the edge of one of the chips, around 8 arcminutes from the aimpoint.
 Finally, there are also some brief pre-2002 observations that we do not include.
\begin{figure}
\begin{center}
\includegraphics[width=0.95\textwidth]{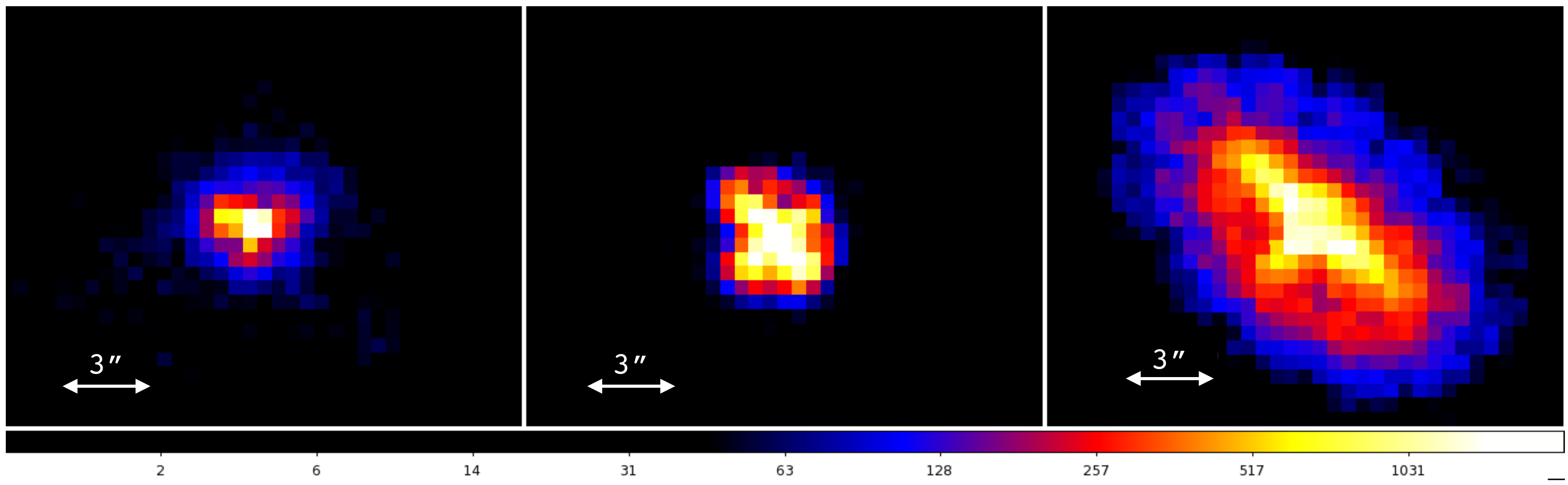}
\end{center}
\caption{NGC1275 in three types of observation, from left to right: centre of the chip in 2004 (\emph{Chandra} ObsID 4952), midway between the edge and centre of the chip in 2009 (ObsID 11714), and the edge of the chip in 2009 (ObsID 11713). The colour coding is adjusted to account for the different observation times such that each colour corresponds to the same count-rate across images.}
\label{fig:ObsPictures}
\end{figure}
\begin{table}[]
\centering
\def\arraystretch{1.2}
\begin{tabular}{r | r | l | c | l}
Obs ID & Exposure {[}ks{]} & Year & Instrument & Location of NGC1275 \\ \hline\hline
3209   & 95.77             & 2002 & ACIS-S     & Central  \\
3404   & 5.31              & 2002 & ACIS-S     & Central  \\
4289   & 95.41             & 2002 & ACIS-S     & Central  \\
4946   & 23.66             & 2004 & ACIS-S     & Central  \\
4947   & 29.79             & 2004 & ACIS-S     & Central  \\
6139   & 56.43             & 2004 & ACIS-S     & Central  \\
6145   & 85                & 2004 & ACIS-S     & Central  \\
4948   & 118.61            & 2004 & ACIS-S     & Central  \\
4949   & 29.38             & 2004 & ACIS-S     & Central  \\
6146   & 47.13             & 2004 & ACIS-S     & Central  \\
4950   & 96.92             & 2004 & ACIS-S     & Central  \\
4951   & 96.12             & 2004 & ACIS-S     & Central  \\
4952   & 164.24            & 2004 & ACIS-S     & Central  \\
4953   & 30.08             & 2004 & ACIS-S     & Central  \\\hline
11715  & 73.36             & 2009 & ACIS-I     & Midway   \\
11716  & 39.64             & 2009 & ACIS-I     & Midway   \\
12037  & 84.63             & 2009 & ACIS-I     & Midway   \\
11714  & 91.99             & 2009 & ACIS-I     & Midway   \\\hline
11713  & 112.24            & 2009 & ACIS-I     & Edge     \\
12025  & 17.93             & 2009 & ACIS-I     & Edge     \\
12033  & 18.89             & 2009 & ACIS-I     & Edge     \\
12036  & 47.92             & 2009 & ACIS-I     & Edge     \\
\end{tabular}
\caption{The Chandra observations used in this paper. The last column shows the location of NGC1275
in the respective observation relative to the focal point.}
% \label{my-label}
\end{table}

The relevance of this classification is that the point spread function of the telescope grows off-axis. In the first group, the photons
from the AGN suffer little dispersion and are highly concentrated on a few pixels. In the third group, the arriving photons are scattered over many pixels,
whereas the second group is intermediate.

The consequence of this is that, despite having the shortest observation time, it is
the third group (the 200ks in which NGC1275 is at the edge of the detector) that provides the cleanest data set. In the first
case, the superb optics of \emph{Chandra} work against it; almost all photons are concentrated onto a few pixels, and these central
pixels are highly contaminated by pile-up.
In this case a clean spectrum can only be obtained by extracting from the wings of the point spread function -- which however reduces the photon count.
In the last case however, being highly off-axis causes
sufficient degradation of the optics that the arriving photons are scattered over many pixels, resulting in greatly reduced pile-up. The second grouping is
intermediary in quality between these two. This is illustrated in Figure~\ref{fig:ObsPictures}, which shows images of NGC1275 for each of the
different observation types.

Another relevant factor is that the brightness of the AGN varies substantially with time. As described in~\cite{Fabian:2015kua}, the NGC1275
AGN was brightest from 1970 to 1990, before rapidly declining by an order of magnitude until around 2000. Since then it appears to have brightened significantly
over the decade from 2003 onwards, although it has not yet returned to the luminosities it had pre-1990.

The combination of pile-up and the intrinsic brightening of the AGN implies that, despite the shorter observation times,
each of the stacked 200ks and 300ks ACIS-I edge and midway observations from 2009 has more counts than the 1Ms of ACIS-S observations taken in 2002 and 2004.
Furthermore, when we consider cleaned spectra that exclude regions of high pile-up,
the total counts in the ACIS-I edge spectra is more than in the ACIS-I midway and ACIS-S observations put together. These factors make the ACIS-I edge observations
the optimal for analysis purposes, even though they involve the shortest observational time.

\subsection{XMM-Newton}

There are two observations of NGC1275 with significant exposure time, see Table~\ref{XMMobs}. The first one, 0085110101, was taken in 2001 when the AGN had its
lowest emissivity in observational history. The second observation, 0305780101, was taken when the emissivity was still relatively low but nevertheless almost twice as bright
as in 2001. NGC1275 is on-axis in both observations.

The 2001 observation was taken in full frame mode, while the 2006 observation was taken in extended full frame mode. This affects the frame
time of the pn camera, which is 73.4 ms in full frame mode and 199.1 ms in extended full frame mode. For MOS, the frame time is 2.6 s. The pixel size of pn is 4.1 arcseconds and 1.1 arcseconds for MOS. This means that pn in extended full frame mode is more susceptible to pile-up than MOS. In general, pile-up is an issue for all \emph{XMM-Newton} observations of NGC1275 as we will discuss in the \emph{XMM-Newton} analysis Section~\ref{XMMAnalysis_sec}.

A significant difference between \emph{XMM-Newton} and \emph{Chandra} is the angular resolution. For both MOS and pn the radius of the disk containing 50 \% of the photons collected in the focal plane (Half Energy Width) is around 8.5 arcseconds at 1.5 keV, while for \emph{Chandra} it is much smaller at $\sim 0.5$ arcseconds.
As the central region of the Perseus cluster is also intrinsically bright, this makes it harder to separate AGN and cluster emission for the \emph{XMM-Newton} observations.

The effective area at 1 keV is 922 cm$^2$ for MOS and 1227 cm$^2$ for pn compared to 340 cm$^2$ for \emph{Chandra}. This allows MOS and pn combined to collect roughly 7 times more photons in a given observation time than \emph{Chandra} (although this also increases the amount of pile-up).
However, there is roughly 10 times more exposure time in the \emph{Chandra} dataset and the significantly better angular resolution of \emph{Chandra} allows a much better contrast to be attained between the AGN and the cluster emission. This leads to much better statistics in the \emph{Chandra} dataset in terms of the total number of counts after background subtraction.

\begin{table}[]
\centering
\def\arraystretch{1.2}
\begin{tabular}{r | r | l | c | l}
Obs ID & Exposure {[}ks{]} & Year & Instrument & Location of NGC1275 \\ \hline\hline
0085110101   & 53.08             & 2001 & EPIC     & Central  \\
0305780101   & 123.3             & 2006 & EPIC     & Central  \\
\end{tabular}
\caption{The \emph{XMM-Newton} observations used in this paper. The last column shows the location of NGC1275 in the respective observation relative to the focal point.}
\label{XMMobs}
\end{table}

\section{Chandra Analysis}
\label{sec:analysis}

We use \texttt{CIAO 4.7}~\cite{ciao}, \texttt{Sherpa}~\cite{sherpa} and \texttt{HEASOFT 6.17} for the \emph{Chandra} data analysis.\footnote{The update to \texttt{CIAO 4.8} affects data taken in Continuous Clocking mode,
 which does not apply to these observations.} After the data is reprocessed using \texttt{CALDB 4.6.9}, it is cleaned from time periods that are polluted by flares using the program \texttt{chips}. We find that only Obs ID 4950 is affected by flares and the cleaning reduces the observation time slightly from 96.12 ks to 89.23 ks for this observation.

\subsection{Pile-Up: General Comments}
\label{sec:pileupcomments}

As the brightness of the AGN makes pile-up a significant feature of all observations, we
will make some general comments on it here.

The energy recorded on the individual ACIS pixels (each approximately 0.5 arcseconds square) is read out approximately every three seconds (one frame time) in these observations. Based on groupings of $3 \ti 3$ pixels, events are graded. Pile-up refers to the arrival of more than one photon in this grouping
within the same readout frame.
This can lead to the energy of the two (or more) incident photons being summed,
and either treated incorrectly as a single photon event of higher energy, or assigned a bad grade (grade migration). For an on-axis bright source (as in the ACIS-S observations of NGC1275), the level of pile-up can be high,
and the resulting spectrum contains events with two, three (and more) photons.
As pile-up is a statistical feature of the number of arriving photons, some
level of pile-up is inevitable in any observation. The question is always whether the magnitude of pile-up is sufficient to corrupt the
science analysis being undertaken.

In terms of the measured photon distribution, the general effect of pile-up is to cause a hardening of the spectrum:  two or more lower-energy photons are
misidentified as a single higher-energy photon. This implies that for a fit of a single power-law to a photon distribution, as pile-up increases
the best-fit power-law index will decrease. In a spectrum contaminated by pile-up, this makes it harder to determine the correct original power-law index.

What about searches for and constraints on ALPs?
As we have seen in Section~\ref{sec:ALPS}, the distinct signal of ALPs is a quasi-sinusoidal modulation in the spectrum -- a local excess or
deficit in the photon count rate
compared to the nearby continuum. For sufficiently small levels of pile-up, localised modulations will remain localised modulations, as an overall
 global continuum redistribution of photons is unable to create or remove localised spikes (or dips) relative to the continuum.\footnote{In the case
 of a strong low-energy emission line, this is not true, as its pile-up may result in spikes at integer multiples of the original line.}
However, for sufficiently heavy pile-up, the majority of photons will be redistributed and such local features will be lost.

While pile-up is always a contaminant on the spectrum,
what this implies is that a search for localised spectral irregularities is more robust against pile-up than, for example, a measurement of the overall power-law spectral index
of a source. While some movement of counts from low to high energies will reduce the number of low-energy photons compared to high energy photons,
it will be less likely to affect the presence or absence of sharp localised features.

This robustness is more applicable at low energies. The effective area of the \emph{Chandra} telescopes starts falling rapidly above around 5~keV.\footnote{cf. \texttt{cxc.harvard.edu/proposer/POG/html/ACIS.html}}
As at higher energies a power-law distribution also produces intrinsically fewer photons, it only requires a small amount of pile-up of lower-energy photons
into the $E > 5 $ keV region to cause a significant distortion of the spectrum there.

In contrast, at lower energies the effective area is larger and there
are far more photons, so small amounts of pile-up will not affect any spectral features. For observations towards Perseus, there is an
additional benefit: the high galactic absorbing column density ($n_H = 1.5 \ti 10^{21} \, {\rm cm}^{-2}$) removes the lowest-energy photons, resulting
in an effective minimal value for a piled-up energy $E_1 + E_2$ (this will be relevant when we discuss in detail below a feature at 2--2.2~keV).

There is an important caveat to this which requires careful treatment. In the presence of rapid variations in the effective area, a failure to
account for pile-up can
result in significant spectral distortion. This is because the distribution of pile-up photons around $E_1 + E_2$ does not respect the behaviour of the effective
area at $E_1 + E_2$, but instead only the effective areas near $E_1$ and $E_2$. If the effective area is rapidly varying near $E_1 + E_2$, and
pile-up photons make up a large fraction of the total photon count at $E_1 + E_2$, then a fitted spectrum will introduce
unphysical features at $E_1 + E_2$.

In the case of \emph{Chandra}, there is a sharp dip in the effective area at around 2 keV arising from an Iridium edge, and the effective area also falls off rapidly
for energies above 5 keV. Between 2.5 and 5 keV the effective area is relatively constant and smoothly varying. With XMM-Newton, there is also significant structure in the effective
area up to around 2.4 keV, but beyond that is smooth and slowly varying up to 7 keV.

The level of pile-up differs greatly between the three observational groups. In the ACIS-S observations, the central pixels are very heavily
affected by pile-up. As an illustration, for an extraction of the ACIS-S spectrum with no exclusion of a central region,
over 5\% of photon counts have energies in the 10-15~keV range. As the effective area of the
telescope is zero at these energies, these counts all arise from piled-up events
in which several photons have arrived in the same readout frame.
In contrast, pile-up is far more moderate for the ACIS-I edge observations
(for which a similar extraction across the entirety of the image results in only 0.1 \% of counts lying in the
10-15~keV range).

Given the data,
there are two basic methods we can employ to reduce the effects of pile-up on the analysis. The first is to reduce the amount
of pile-up by using an annular extraction region and extracting the spectrum only
from the wings of the point spread function. The point spread function of the telescope causes the arriving photons to be spread out
on the detector, with the degree of spread increasing as one moves progressively off-axis. Furthermore, the point spread function is mildly energy dependent: photons with higher energy are spread out further than photons with smaller energy. While fewer photons arrive in the wings of the point spread function,
those that do suffer less from pile-up than those arriving on the central pixels.
The second is to model the pile-up, and the distortion it induces on the spectrum, explicitly. We shall utilise both approaches %%Sections~\ref{subsec:Clean} and~\ref{subsec:Pileup}
below.

The overall balance here is between enhanced photon statistics -- but with more pile-up and so
with worse data quality -- and fewer photons but better data. In particle physics language, this is the trade-off between efficiency and purity.
%%We however start with analyses involving complete extractions of the spectra, with no exclusions of high pile-up regions.

\subsection{Fitting Accuracy and Bounds}

Before we describe the data in detail, we comment on the relationship between the accuracy of the fit and the
bounds on ALP parameters that we are able to produce. This is relevant as the data involves a bright point source at the middle
of a bright and complex continuum background (the centre of the Perseus cluster).

Constraints on ALPs arise because they induce modulations that would make the conventional fit (an absorbed power law) a bad one.
The constraints on ALPs come from requiring that the badness-of-fit that ALPs would induce is no worse than that present in the actual data --
a coupling $g_{a\gamma\gamma}$ is excluded when it would lead to a fit that is worse than actually occurs. To place bounds on ALPs in this way,
it is not necessary that the actual data be a perfect fit. It is true that the better the fit of the actual data to an absorbed power law, the stronger are the bounds that can be placed on ALPs. However, even if the actual data is only -- for example -- within 10\% of the standard fit, then this still implies the exclusion of ALP couplings large enough to give 30\% modulations in the data.

This method of producing bounds is conservative. The simulated data we use when considering the effects of ALPs is more idealised than for real data. In particular, the simulated data does not account for any possible observational or instrumental inaccuracies arising from the complexity for the source, such as pileup from a high photon flux or possible mis-subtraction of the cluster background when obtaining the AGN spectrum.

On the other hand, in the actual data there are various observational, instrumental or astrophysical effects that could cause the standard fit to fail to be a perfect description of the measured data, even without any exotic physics.
These include
\begin{enumerate}
\item
As mentioned above, pile-up can lead to photons being registered at the wrong energy. This can lead to fake excesses at the location
of effective area dips.
\item
The X-ray emission at the centre of the Perseus cluster has an intensity that varies spatially away from the nucleus. As the background regions
differ from the source region used, the process of background subtraction may lead to residual cluster emission being present
in addition to the AGN spectrum.
\item
The response files may not deal fully accurately with cases
where the centre of the image has significant pile-up.
\end{enumerate}

We will exclude ALP couplings for which idealised simulated data is still a worse fit compared to the noisier actual data.
`Noise' in the actual data reduces the bounds that can be placed on ALPs (as small genuine ALP-induced modulations could hide
under the noise), but still robustly allows the exclusion of large ALP-induced modulations.
The bounds that we place here are therefore conservative, but with the potential of further improvement
from better modelling of the actual data or through observing NGC1275 with cleaner settings (such as a shorter read-out time).

\subsection{Spectral Analysis}

We first extract %%complete
the spectrum for NGC1275 for the ACIS-I edge observations (11713, 12025, 12033, 12036), where the AGN is around seven arcminutes off the optical axis, using the full extraction region without any exclusion of the central core. This is only valid for the edge observations
where pile-up is relatively low throughout the whole image.
Spectra and responses were created using \texttt{specextract} for each observation, and then
stacked using \texttt{combine\_spectra}. An ellipse around NGC1275 of radii 11.6 and 7.2 arcseconds was used for the extraction region.
The boundaries of the ellipse were set by the location where the image of the AGN ceased to dominate the background cluster emission.
The background was taken from an elliptical annulus around NGC1275, with the outer radii being 19 and 13 arcseconds and the inner
 radii 13.3 and 9.3 arcseconds (for the two short observations 12025 and 12033 this region goes beyond the edge of the
 chip, and a rectangular box was used instead for the background).

The resulting stacked spectrum contains around 266000 counts, reducing to 230000 after background subtraction, giving a ratio of 6.5:1 for the AGN against the cluster emision.
%After background subtraction, the fraction of counts in the 7--10~keV band and 10--15~keV bands were 1.3 \% and 0.09 \% respectively (we use these percentiles as a loose
%measure of the extent of pile-up contamination).
The resulting spectrum was binned to ensure a
minimum of 2000 counts a bin,\footnote{In general we bin so that there are approximately
one hundred bins in total. If there are too few counts per bin, then the fit is insensitive to
localised modulations as can be produced by ALPs, as the goodness-of-fit is insensitive to the sign of the residuals.} and fitted between 0.8 and 5~keV with
an absorbed power law \texttt{xswabs} $\ti$ \texttt{powlaw1d},
\begin{equation}
A E^ {- \gamma} \times e^{-n_{H} \sigma(E)}\, .
\label{eq:absorbedpowerlaw}
\end{equation}
Here $A$ denotes the normalisation of the power-law, $\gamma$ the power-law index, and $n_H$ the effective Hydrogen column density.

\begin{figure}
\begin{center}
\includegraphics[width=0.85\textwidth]{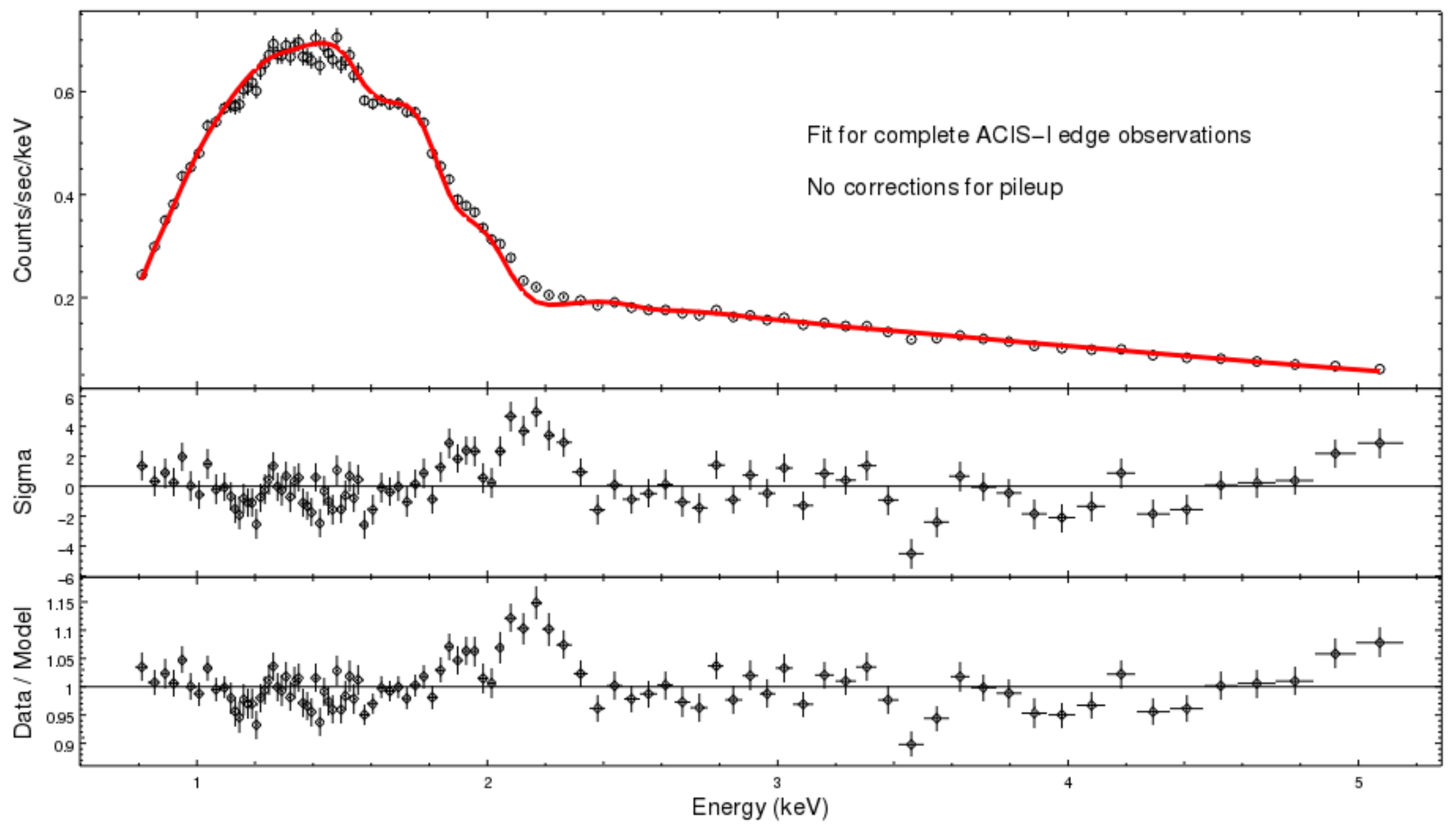}
\end{center}
\caption{The stacked spectrum of the ACIS-I edge observations from the entirety of the extraction region,
involving 230000 counts after background subtraction.
 The fit is to an absorbed power law, and results in $n_H = 2.3 \ti 10^{21} {\rm cm}^{-2}$ and a spectral index
of $\gamma=1.83$. $\sigma$ refers to the standard deviation from the model expectation for a Poissonian count rate.}
\label{fig:ACISIEdgeFull}
\end{figure}

The resulting fit is shown in Figure~\ref{fig:ACISIEdgeFull}, together with the fractional ratio of data to model.
 The best-fit value of $n_H$ is $2.3 \ti 10^{21} {\rm cm}^{-2}$ and the power-law index is
$\gamma = 1.83 \pm 0.01$.  While the absorbed power-law is a reasonable characterisation of the data,
there are two large
localised residuals: one positive between 2--2.2~keV and one negative around 3.4--3.6~keV. There is an upward trend at 5~keV. As the effective area of \emph{Chandra}
begins to fall off rapidly here, and there are also intrinsically fewer photons expected, pile-up plays a proportionately more important role.
This rising trend continues beyond 5 keV and we attribute this to the effects of pile-up.

For the 300ks of ACIS-I Midway observations with the AGN around 3 arcminutes from the aimpoint, a comparable
full extraction results in a spectrum that is still highly piled up (the
best-fit index is an unphysically hard $\gamma = 1.30$). This is even more the case for the ACIS-S observations, where a similar full extraction produces an index of
$\gamma = 0.66$. For these cases, it is necessary to account for pile-up to extract physical results.

Given the high flux levels of the source, a cleaner spectrum requires reducing the effects of pile-up in the spectrum.
%%In this section we
%%describe
There are two ways of reducing the effects of pile-up. The first method involves excluding a central region of highest pile-up from the analysis. While this reduces the photon counts -- and so the statistical significance of any features in the spectrum -- it also allows produces
purer spectra of higher quality. The second method is to keep the full spectrum, but model the pile-up explicitly.
For the purpose of placing bounds on ALPs and constraining the existence of ALP-induced spectral modulations,
 the most useful spectra are those where photon events occur at their true energies.
We therefore only describe in the main text the creation and analysis of spectra with a central region of
high pile-up excluded. In Appendix~\ref{subsec:Pileup} we also discuss the explicit modelling of pileup in cases where we retain the entirety of the image.

We analyse the images and count rates
of each observation using the image software \texttt{ds9}.
The precise method of central exclusion is somewhat arbitrary; the method used to produce the spectra was as follows.
We set the \texttt{ds9} binning such that
the \texttt{ds9} pixels are essentially the same size as the physical \emph{Chandra} pixels
(half an arc-second across).
We then create an extraction region manually excluding all \texttt{ds9} pixels which are neighbours (either side-by-side or diagonally) to any pixel with total counts
greater than 1\% of the overall exposure time. Using the \texttt{pileup\_map} tool, this results in almost all retained pixels having a pile-up fraction lower
than 5\%.\footnote{For these observations, the telemetry limit was 15 keV and any (piled-up) photons recorded with energy larger than this were not telemetered to the ground. For there to be a significant number of such events, there would also exist a tail of photons with energies
close to, but below, the telemetry limit. From the $< 0.2\%$ fraction of photons measured in the 10-15 keV band for the cleaned spectra, we can infer that these contained very few un-telemetered events. The same analysis shows that there are very few un-telemetered events for the
ACIS-I edge observations even without excising the central part of the image, but the central high pile-up regions of the ACIS-I Midway and ACIS-S observations did have significant numbers of un-telemetered events.} Using {\texttt arfcorr} the responses reflect the presence of a central
exclusion in the extraction region.

\begin{figure}
\begin{center}
\includegraphics[width=0.8\textwidth]{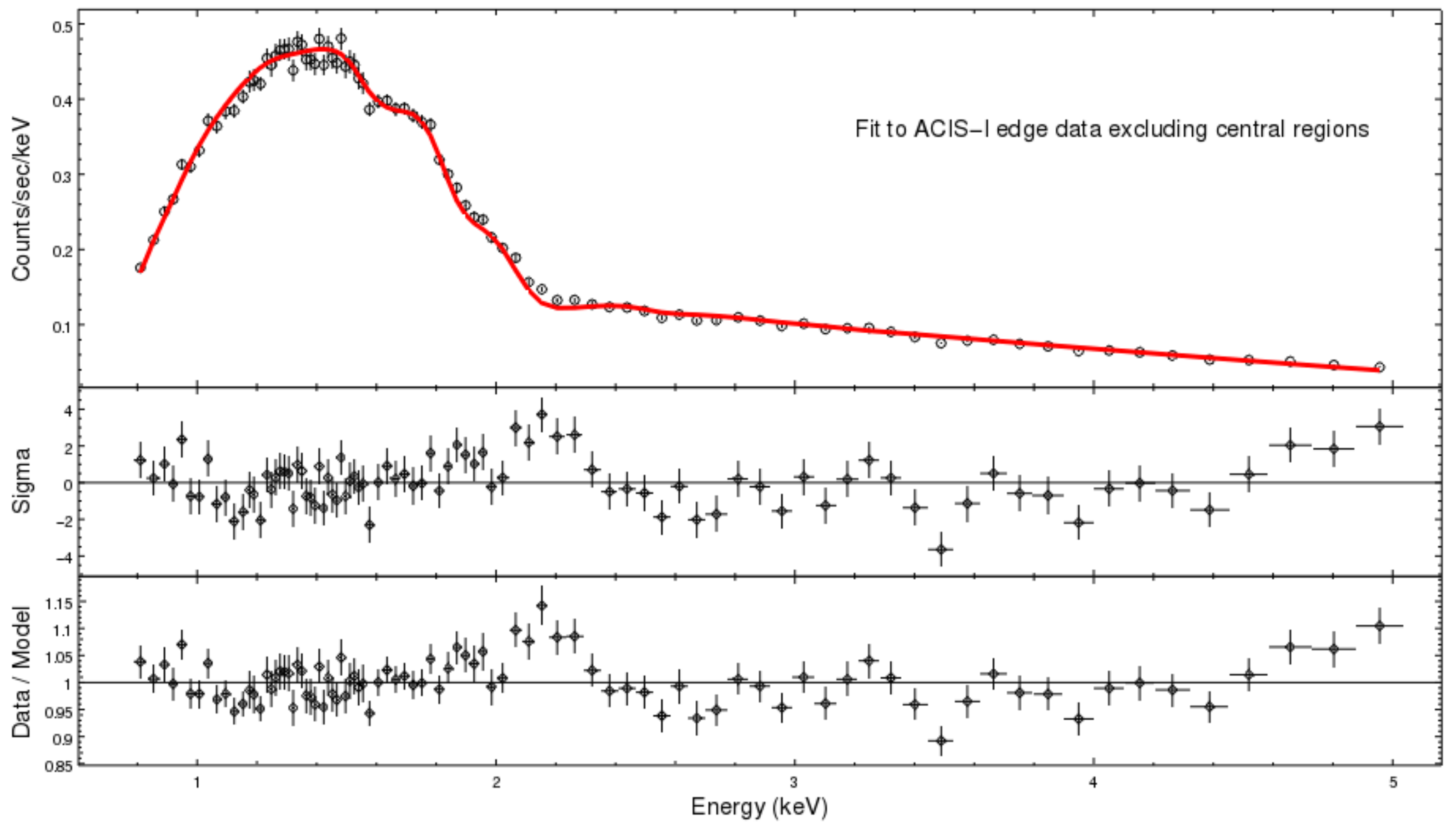}
\end{center}
\caption{The stacked spectrum of the ACIS-I edge observations, with central pixels excluded according to the prescription in the text.
There are 153000 counts after background subtraction.  The fit is to an absorbed power law, and results in $n_H = 2.2 \ti 10^{21}{\rm cm}^{-2}$ and a spectral index
of $\gamma=1.85 \pm 0.015$.}
\label{fig:ACISIEdgeClean}
\end{figure}

We first show in Figure \ref{fig:ACISIEdgeClean} the ACIS-I edge observations with a central exclusion according to the above method.
As these observations are relatively clean to begin with, the fractional change
in the number of counts is relatively small. There are now 187000 counts before background subtraction and 153000 counts afterwards.
The fraction of counts in the 7--10~keV and 10--15~keV regions are 1.2\% and 0.04 \%. The best-fit spectral index is $\gamma=1.85 \pm 0.015$ (the errors quoted here and in the remainder of this paper are statistical 1-$\sigma$ errors, see Appendix~\ref{sec:appb3} for more details) with $n_H = 2.2 \ti 10^{21} {\rm cm}^{-2}$.
While the statistical significance of the features at 2--2.2~keV and 3.4--3.5~keV reduces (consistent with the reduced photon counts),
the magnitudes of the data to model fluctuations remain the same.

\begin{figure}
\begin{center}
\includegraphics[width=0.8\textwidth]{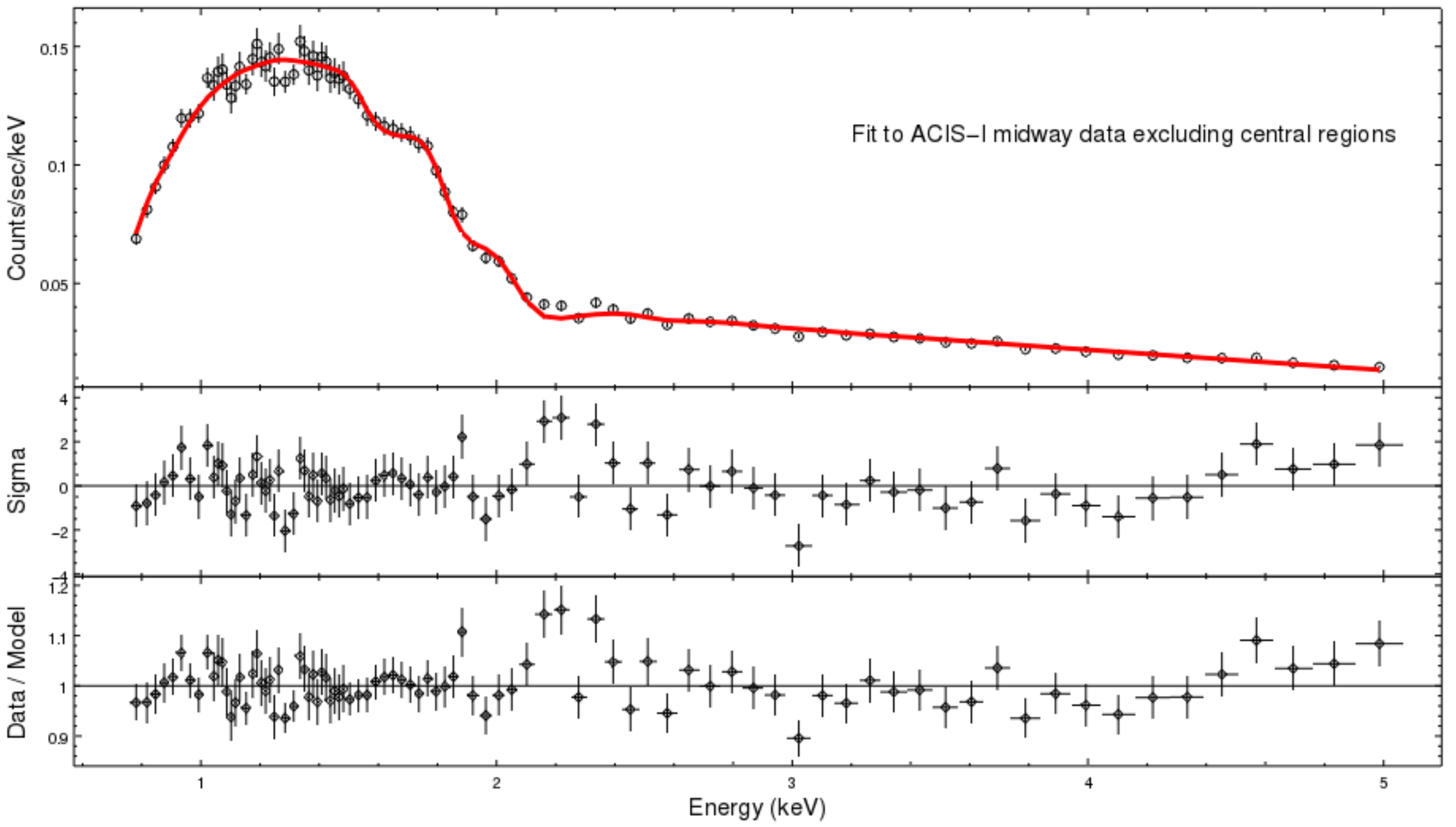}
\end{center}
\caption{The stacked spectrum of the ACIS-I midway observations, with central pixels excluded according to the prescription in the text.
There are 74000 counts after background subtraction.  The fit is to an absorbed power law, and results in $n_H = 1.3 \ti 10^{21}{\rm cm}^{-2}$ and a spectral index
of $\gamma=1.64 \pm 0.02$.}
\label{fig:ACISICentralClean}
\end{figure}

We now perform a similar cleaning of the ACIS-I observations with the source midway on the chip.
As the central pixels are heavily piled-up, in this case the cleaning procedures significantly increases
the quality of the fit. This is at the cost of a significant reduction in photon statistics: there are now only
88000 counts before background subtraction, and 74000 after background subtraction.
We group counts so that there are 700 counts per bin, leading to
a spectral index of $\gamma=1.64 \pm 0.02$ with $n_H = 1.3 \ti 10^{21} {\rm cm}^{-2}$. With a Q-value of 0.18 and a reduced $\chi^2$ of 1.14 for 86 degrees of freedoms,
this is now an overall
good fit to the data (see Figure~\ref{fig:ACISICentralClean}).\footnote{Fitting the different observations individually produces results for $\gamma$ that are consistent with the stacked result, both for the Midway and Edge observations.}
%The cleaned spectrum still shows a clear excess in the 2.1--2.2~keV region at a ratio consistent with other observations.
%While not at notable significance, there is also a mild local deficit at 3.4--3.5~keV.
%In both cases, the ratio of these features in terms of data to model is entirely consistent with the high-statistics results from the full
%ACIS-I edge observations.

%% JC edited para
We note that the spectral index $\gamma$ is clearly different from that found with the ACIS-I edge data. However, this spectrum
was obtained by excluding the central part of the image and extracting only from the wings of the point spread function.
By comparison with the brightness of the AGN measured in the contemporaneous ACIS-I edge observations, we can infer that
this spectrum retains only around one quarter of the total number of photons from the AGN.
As the ACIS point spread function increases with energy, this biases the analysed photons towards a harder spectrum (and a lower $\gamma$).
In principle, this is taken into account in the analysis procedure, as the aperture correction of {\texttt arfcorr} modifies the response to
reflect the exclusion of the central part of the image. However, it is also the case that the image has a
 complex form with neither a spherical nor an elliptical shape. Instead, the image of the AGN -- traced by the excluded central region -- resembles a `Maltese cross' shape, as can be seen in Figure 2.

 The objective of this work is to constrain large, localised modulations in the spectrum of the AGN that could arise from ALP-photon
 conversion. For this purpose
 we therefore do not regard these different values of $\gamma$ as problematic, particularly given the large difference in observational conditions between the two sets of observations and the large fraction of photons that are excluded in the ACIS-I Midway observations.

\begin{figure}
\begin{center}
\includegraphics[width=0.8\textwidth]{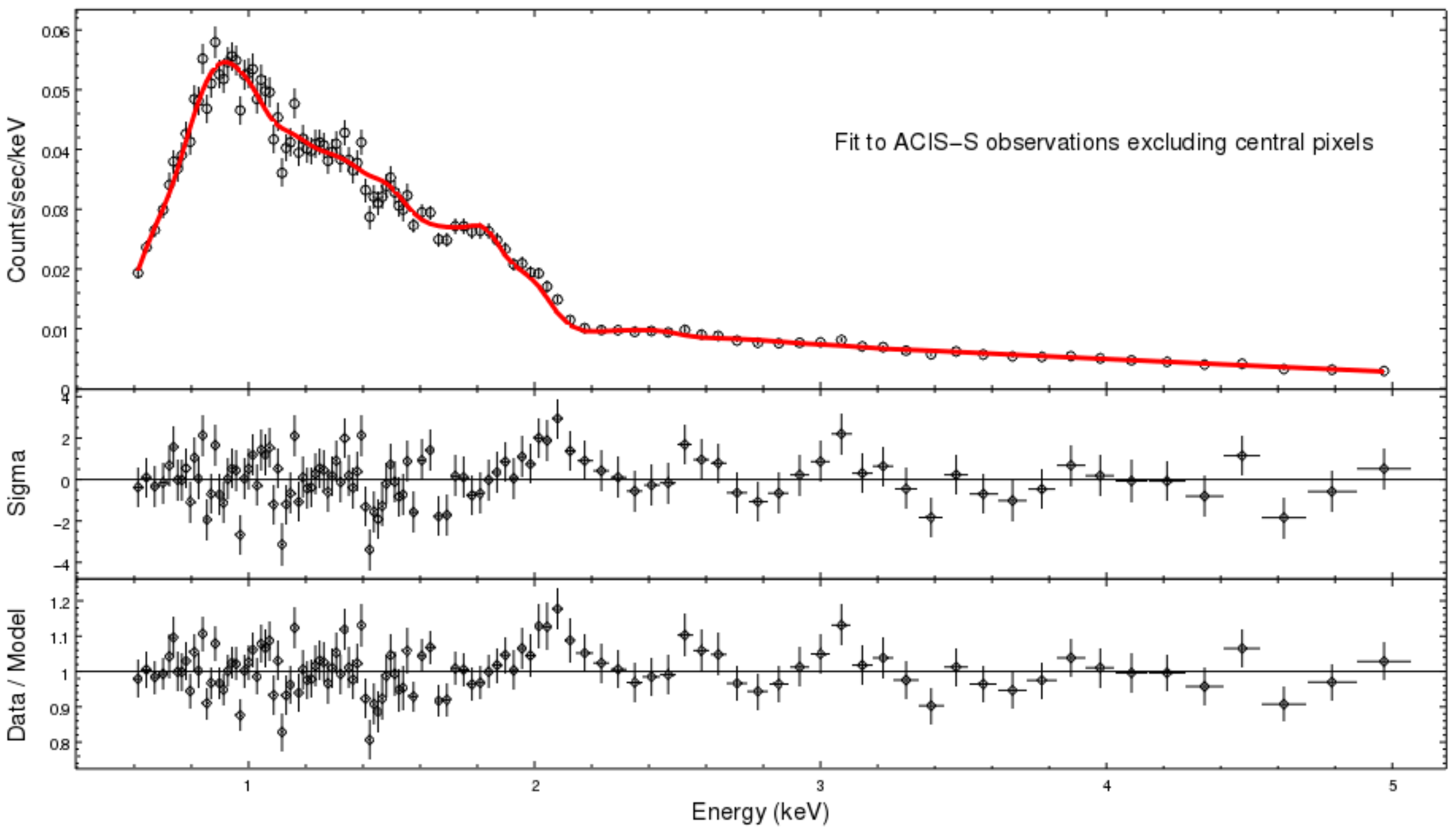}
\end{center}
\caption{The stacked spectrum of the ACIS-S observations, with a central 16-pixel square excluded.
The fit is to an absorbed power law and soft thermal component.}
\label{fig:ACISSClean}
\end{figure}

We perform a similar procedure for the ACIS-S observations. In this case we reduce pile-up by removing a central
square of 16 pixels (2 arcseconds square) from an extraction region of a circle of radius 3.4 arcseconds. The resulting cleaned spectrum has 117000 counts before
background subtraction and 74000 counts after background subtraction. After background subtraction
there are now 0.1\% counts in the 10--15~keV band and 1.1\% of counts in the 7--10~keV band, indicating that this spectrum is now substantially cleaner.
We group counts so that there are at least 700 counts per bin. In this case, an absorbed power-law is not sufficient for a good fit and we supplement this by
a soft thermal component using \texttt{xsapec} (the presence of a thermal component for NGC1275 was also reported in~\cite{Yamazaki}).
The presence of a soft thermal component substantially improves the fit, resulting in an acceptable Q value of $10^{-2}$.
The resulting fit (see Figure~\ref{fig:ACISSClean}) has $n_H = 1.6 \ti 10^{21} {\rm cm}^{-2}$ and a power-law index of $\gamma=1.84 \pm 0.03$. The thermal component has a
temperature $T = 0.92 {\rm keV}$. At this temperature, the amplitude and abundance of the thermal component are largely degenerate in the fit.
Fixing the abundance at solar abundance, the relative amplitude of thermal component to the power-law is 0.15.

Why is a soft thermal component necessary for a good fit in the ACIS-S observations but not in the ACIS-I observations?
There are two main reasons.
First, the effective area of ACIS-S has more support at low energies than for ACIS-I. The net result is that for ACIS-S
the peak of the observed photon spectrum is around 1~keV, whereas for the ACIS-I observations the peak is around 1.6~keV.
This implies that the ACIS-S observations have a proportionally higher fraction of low energy photons, and so are
more sensitive to the presence of a soft thermal component.

Secondly, the ACIS-S and ACIS-I observations were taken at different times - the former in 2002 and 2004 and the latter in 2009. Over this period the intrinsic brightness of the AGN, as measured by the power-law component, has increased significantly.
Assuming that the soft thermal component arises differently from the hard power-law, and that the
absolute strength of the soft thermal component has remained the same over this period, its relative
significance would be much greater in the ACIS-S observations.

We have checked that one can add a soft thermal component to the ACIS-I spectra, of the same absolute magnitude as for the ACIS-S spectra, without affecting the quality of the fit.
As for the ACIS-I observations the thermal component neither significantly improves or worsens the fit, and does not affect the above results, we have not included it explicitly
for the ACIS-I data.

Finally, we also considered \emph{Chandra}'s grating observations of NGC1275 (ObsIDs 333
and 428 with HETG) as a means to reduce pile-up, but this reduced counts to the point
where it did not seem worthwhile.

\subsection{Summary of Fits for ALP Constraints}

In terms of constraining ALPs and their parameter space, the main consistent feature observed from these plots
is that there are no residuals seen above a level of $10 \%$. This result is found across all the three distinct
classes of observational conditions. If they exist with strong enough coupling, ALPs can induce localised fractional modulations of up to 50\% in the spectrum. It is unambiguously the case that no modulations of this size are present in the spectrum, and therefore
ALP-photon couplings large enough to generate such modulations are excluded.
As any mis-modelling of the actual data would worsen the fit to the real data, and \emph{increase} the size of residuals in the real data,
this ability to constrain ALP parameter space is robust against
any such mis-modelling that may be present.
Before we determine constraints on ALPs, however, for completeness we describe the XMM-Newton observations of NGC1275.

\section{XMM-Newton Analysis}  \label{XMMAnalysis_sec}

We use \texttt{SAS} version 15.0.0. The 2001 and 2006 datasets are reprocessed with up to date callibrations via \texttt{cifbuild}. To remove flares we apply the standard filters of counts/s $< 0.35$ for MOS and counts/s $< 0.4$ for pn. There is a significant flare towards the end of the 2006 observational period and various shorter flares during the 2001 observation. Removing the polluted time intervals results in reducing the effective exposure time from 53 ks to 49 ks for MOS 2001, 25 ks to 7.4 ks for pn 2001, 123 ks to 117 ks for MOS 2006 and 76 ks to 49 ks for pn 2006.

For the extraction regions we first choose circle of radii 13.8 arcseconds (MOS1 2001), 17.5'' (MOS2 2001), 14.3'' (MOS1 2006), 17.5'' (MOS2 2006), 15.5'' (pn 2001), and 19'' (pn 2006). We then check if pile-up is an issue using the \texttt{SAS} tool \texttt{epatplot} and find that pile-up is in general present in all the observations. Since the AGN was less bright by a factor $\sim$ 2 in 2001, pile-up is slightly less of an issue for these observations. On the one hand, the smaller pixel size of 1.1 arcseconds of the MOS cameras makes it less susceptible for pile-up than the 4.1 arcseconds pixel size of the pn camera. On the other hand, pn's time resolution of 73.4 ms in full frame mode is advantegous with respect to pile-up compared to the 2.6 s time resolution of MOS. However, in extended full frame mode pn's 199.1 ms time resolution make it less advantegous with respect to pile-up compared to MOS. Note that after flare removal there are only 7.4 ks of pn data in full frame mode while there are 49 ks of
more piled up data in extended
full frame mode. As the MOS data is expected to be more sensitive,
we do not present the pn data from either 2001 or 2006
and instead focus on the MOS data only.

In order to make a compromise between avoiding the most heavily piled-up regions and not too small signal over background ratio we choose the following inner annuli radii 2.5 arcseconds (MOS1 2001), 3'' (MOS2 2001), 7.5'' (MOS1 2006), 7.5'' (MOS2 2006). For the background regions we choose annuli of 13.8-22 arcseconds (MOS1 2001), 17.5-22.5'' (MOS2 2001), 14.3-22.5'' (MOS1 2006), 17.5-22.5'' (MOS2 2006). The worse angular resolution of \emph{XMM-Newton} compared to \emph{Chandra} implies that the discrimination
 of the AGN emission against the cluster is relatively poor (for the 2006 data, the AGN:cluster data ratio is 1:3, whereas for \emph{Chandra} spectra this ratio is not worse than
 1.7:1).

After applying \texttt{eveselect} to generate the spectra we use \texttt{rmfgen} and \texttt{arfgen} to generate the response files. We use \texttt{epicspeccombine} to combine the MOS1 and MOS2 spectra of the 2001 and 2006 obervations, respectively. We also use the ftools routines \texttt{mathpha}, \texttt{addrmf} and \texttt{addarf} (\texttt{epicspeccombine} does not produce an arf file) and find that the results are comparable within statistical errors to using \texttt{epicspeccombine}. In order to avoid combining different systematic errors we do not combine the 2001 and 2006 data.

The spectra are further analysed using the spectral fitting programs \texttt{Xspec}~\cite{Arnaud} and \texttt{Sherpa}~\cite{sherpa}, requiring a minimum count-rate per bin of about 1.5\% of the overall count rate. The energy interval is 0.5--7.5~keV.

For NGC1275, we use a spectral model of a powerlaw as in our Chandra analysis
\begin{equation}
 A(E) = K\,E^{-\gamma}\,,\label{zpowerlw}
\end{equation}
where $K$ is the normalisation in photons keV$^{-1}$ cm$^{-2}$ s$^{-1}$ and $\gamma$ is the photon index. To model a soft excess of the AGN, we add an \texttt{apec} model whose temperature we expect to be around 1 keV. The abundance parameter of this model is set to unity.\footnote{Note that around 1 keV this \texttt{apec} spectrum is dominated by atomic lines. Hence, there is a degeneracy between the abundance parameter and the normalisation of the \texttt{apec} model.} We also include the well known iron line at 6.4 keV in the cluster rest frame in our spectral fit.

To model photoelectric absorption, the source spectrum is multiplied by the \texttt{zwabs} model
\begin{equation}
 M(E) = e^{-n_H \sigma\left(E\right)}\,,
\end{equation}
where $n_H$ is the hydrogen column density and $\sigma$ is the photoelectric cross-section~\cite{Morrison:1983hg}. Photoelectric absorption is mostly relevant below approximately 1 keV if $n_H \gtrsim 10^{21}$ atoms cm$^{-2}$ (The galactic column density is measured as $n_H = 1.5 \cdot 10^{21}$ atoms cm$^{-2}$.)

\begin{figure}[h!]
	\begin{center}
	\includegraphics[width=0.8\textwidth]{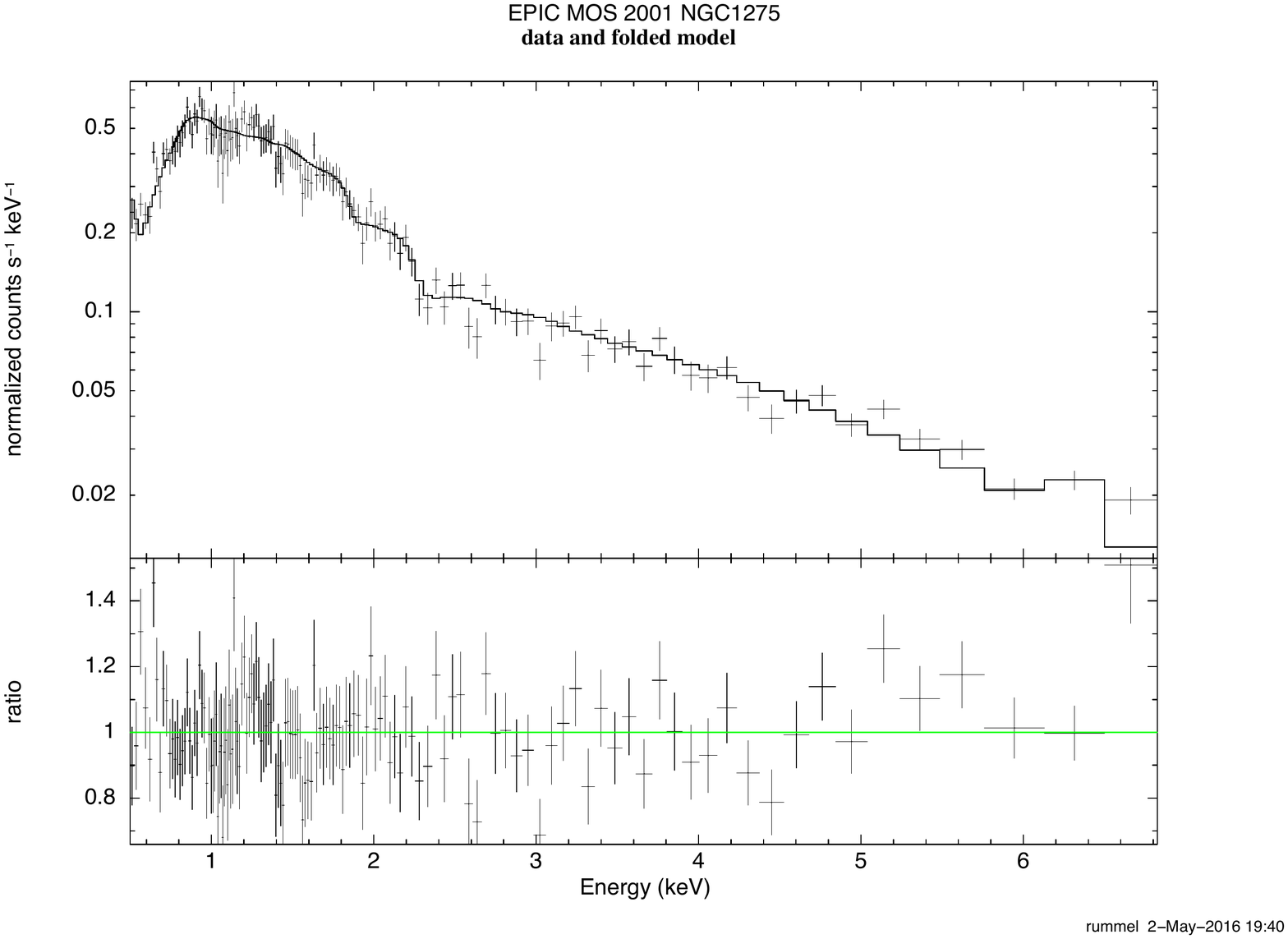}
   \end{center}
 \caption{MOS 2001 spectral fit. The best fit parameters are listed in Table~\ref{fit_tab}.}
 \label{MOS2001_fig}
\end{figure}

\begin{figure}[h!]
	\begin{center}
	\includegraphics[width=0.8\textwidth]{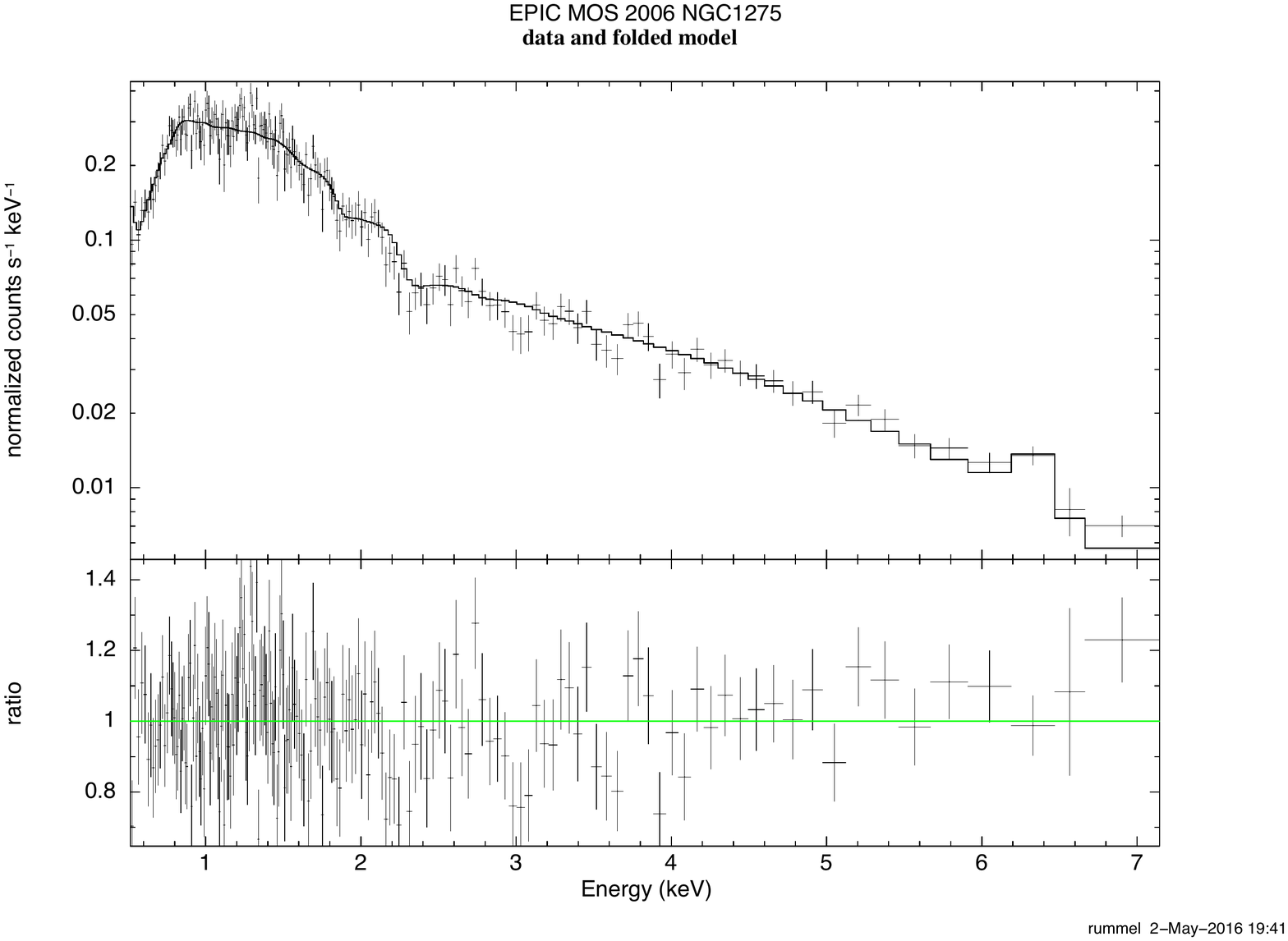}
   \end{center}
 \caption{MOS 2006 spectral fit. The best fit parameters are listed in Table~\ref{fit_tab}.}
 \label{MOS2006_fig}
\end{figure}

%We do not observe any significant excess at 2--2.2~keV or a deficit at 3.5 keV, although there is a mild dip in the 2006 MOS data at 3.5 keV, see Figure~\ref{MOS2006_fig}. However, this is not a contradiction with having these features real and present in the \emph{Chandra} data.
%First of all, the statistics for \emph{XMM-Newton} are more limited: the two spectra, combined, contain a total of $\sim$ 110000 counts for the MOS observations after background subtraction. Secondly, the data is much noisier: the overall spectrum is background dominated, making it harder to ensure
% sensitivity to small fractional residuals in the AGN spectrum.
 %Adding a negative Gaussian feature that corresponds to a 3\% fractional dip in the 3.3--3.7~keV region, see Figure~\ref{RadialProfile}, the goodness of the fit reduces by $\Delta \chi^2 < %Finally, pile-up can significantly alter the spectrum of the \emph{XMM-Newton} observations around 2 -- 2.4 keV,
 %as we now discuss in Section~\ref{XMMpilup_sec}.
The fitted spectra are shown in Figures \ref{MOS2001_fig} and \ref{MOS2006_fig}. While the fit is good (see Table~\ref{fit_tab}), there are both fewer counts and a worse contrast against the cluster background than for the \emph{Chandra} data. From the perspective of ALP constraints, we can say that modulations of the spectrum are not allowed beyond the 20\% level -- a weaker constraint than
for the \emph{Chandra} data.

\begin{table}[]
\centering
\def\arraystretch{1.2}
\begin{tabular}{c | c | c}
 & MOS 2001 & MOS 2006 \\ \hline\hline
Exposure {[}ks{]} & 49 & 118 \\
Counts & $\sim$ 47000 & $\sim$ 63000\\ \hline
$\gamma$ & 1.65 $\pm$ 0.03 & 1.65 $\pm$ 0.03 \\
$T_{\text{soft}}$ & 0.84 $\pm$ 0.06 & 0.80 $\pm$ 0.06 \\
$n_H$ {[} $10^{22}$ atoms cm$^{-2}$ {]} & 0.13 $\pm$ 0.03 & 0.13 $\pm$ 0.03 \\ \hline
$\chi^2 /$ dof & 154 / 131 & 205 / 178 \\
Q-value & 0.08 & 0.08 \\
\end{tabular}
\caption{Fit results for the \emph{XMM-Newton} MOS observations from 2001 and 2006. These results are in agreement with the fit parameters found in~\cite{Churazov:2003hr} for the 2001 MOS data of NGC1275.}
\label{fit_tab}
\end{table}

\section{Bounds on ALPs}
\label{sec:bounds}

From the perspective of constraints on ALPs, the overall summary of the data is that the spectrum from NGC1275 is fit by an
absorbed power-law, with residuals from the power-law not larger than 10\%.  We will further discuss residuals at the 10\% level in Appendix~\ref{sec:signal}
(in particular features around 2 - 2.2 keV and at 3.5 keV).
However, the absence of any modulations in the spectrum at a level greater than 10\% allows powerful constraints to be placed on ALP parameter space. %%CHANGE

In particular, this allows us to say that on passing from NGC1275 through the Perseus cluster and towards us, $\langle P(\gamma \to a) \rangle \lesssim 20 \% $.
It follows that ALP-photon couplings large enough to generate the saturated limit of $\langle P(\gamma \to a) \rangle = 1/3$ are robustly excluded, as these would
produce larger residuals in the data than are actually observed.

Note that this approach to placing bounds is conservative. The simulated data (with ALPs) is idealised -- it has no pileup present and assumes
a perfect subtraction of the cluster background. Couplings are excluded when
they lead to simulated data that would \emph{still} be a worse fit than the actual data, even though the simulated
data is observationally ideal while the real data contains effects such as pileup and cluster background that lead to increased
residuals and make it far noisier than simulated data.

To obtain approximate bounds on $g_{a \gamma \gamma}$, we compare two models for the flux $F(E)$ observed from NGC1275:

\begin{itemize}
\item Model 0: An absorbed power law $F_{0}(E) = A E^ {- \gamma} \times e^{-n_{H} \sigma(E)}$, as described in Equation~\eqref{eq:absorbedpowerlaw}.

\item Model 1: An absorbed power law multiplied by the photon survival probability assuming the existence of ALPs with coupling $g_{a \gamma \gamma}$. In this case the predicted flux also depends upon the magnetic field $ {\bf B}$ along the line of sight. We have  $F_{1}(E, {\bf B}) = A E^ {- \gamma} \times e^{-n_{H} \sigma(E)} \times P_{\gamma \to \gamma} (E (1 + z), { \bf B}, g_{a \gamma \gamma})$.

\end{itemize}

Although we have (limited) empirical estimates of the strength of the magnetic field in Perseus, the exact structure is unknown. In practice, we randomly generate many instances of the field from a given power spectrum. The parameter most relevant to ALP-photon conversion is the central magnetic field strength $B_0$, estimated as $25 \, \mu {\rm G}$ in \cite{0602622}.
 Based on results for the Coma cluster, we assume that $B$ decreases with radius as $B \propto n_{e}^{0.7}$ \cite{1002.0594}. The electron density $n_{e}$ has the radial distribution found in \cite{Churazov:2003hr}, $$n_{e} (r) = \frac{3.9 \times 10^{-2}}{ [ 1 + (\frac{r}{80 \, {\rm kpc}})^2]^{1.8}} +  \frac{4.05 \times 10^{-3}}{ [ 1 + (\frac{r}{280 \, {\rm kpc}})^2]^{0.87}} \, {\rm cm}^{-3}~.$$ We simulate each field realisation with 600 domains. The length $l$ of each domain is between $3.5$ and $10 \, {\rm kpc}$, randomly drawn from a power law distribution with minimum length $3.5 \, {\rm kpc}$ and power $0.8$. We therefore have:

\begin{equation}
P(l = x) = N
\begin{cases}
0 & {\rm for} \, x > 10 \, {\rm kpc}~,\\
 x^{-1.2} \, &{\rm for} \,  3.5 \, {\rm kpc} < x < 10 \, {\rm kpc}~,\\
0  \, & {\rm for} \, x < 3.5 \, {\rm kpc}~,
\end{cases}
\end{equation}
with normalisation constant $N$.

The coherence length and power spectrum of the magnetic field in the centre of Perseus is not observationally determined. Instead, these parameters are motivated by those found for the cool core cluster A2199 \cite{Vacca:2012up}, taking a conservative value for the magnetic field radial scaling.  The magnetic field and electron density are constant in each domain, with $B(r)$ and $n_{e}(r)$ evaluated at the centre of the domain and the direction of ${\bf B}$ chosen at random. \\

We compute $95 \%$ confidence limits on $g_{a \gamma \gamma}$ by generating fake data from Model 1 and assessing how well it is fit by Model 0 i.e.~how well the oscillations due to ALP-photon conversion can hide in the Poisson noise. As it has minimal pile-up, we use the clean ACIS-I edge observations for this analysis. We fit the spectrum between $1$ and $4 \, {\rm keV}$ (a region unaffected by pile-up) and bin such that there are 1000 counts in each energy bin. We use \texttt{Sherpa}'s Levenberg-Marquardt fitting method with Poisson errors derived from the value of the data in each bin. Our procedure to determine whether ALPs with coupling $g_{a \gamma \gamma}$ are excluded at the $95 \%$ confidence level is as follows:

\begin{enumerate}

\item Fit Model 0 to the real data and find the corresponding reduced $\chi^{2}$, $\chi^{2}_{\rm data}.$
\item Randomly generate 50 different magnetic field realisations ${\bf B}_{i}$ for the line of sight to NGC1275.
\item For each ${\bf B}_{i}$, compute $P_{\gamma \to \gamma} (E,  {\bf B}_{i},  g_{a \gamma \gamma})$ by numerically propagating photons at different energies through ${\bf B}_i$, as described for example in \cite{1312.3947}. We take 300 photon energies equally spaced between $1$ and $4 \, {\rm keV}$.
\item  For each ${\bf B}_{i}$, generate 10 fake data sets from Model 1, using \texttt{Sherpa}'s fake pha method.
\item Fit Model 0 to each of the 500 fake data sets and find the corresponding reduced $\chi^{2}$, $\chi^{2}_{i}$ for each.
\item If fewer than $5 \%$ of the $\chi^{2}_{i}$ are lower than $\chi^{2}_{\rm data}$, $g_{a \gamma \gamma}$ is excluded at the $95 \%$ confidence level.

\end{enumerate}

\begin{figure}
\begin{center}
\includegraphics[width=0.8\textwidth]{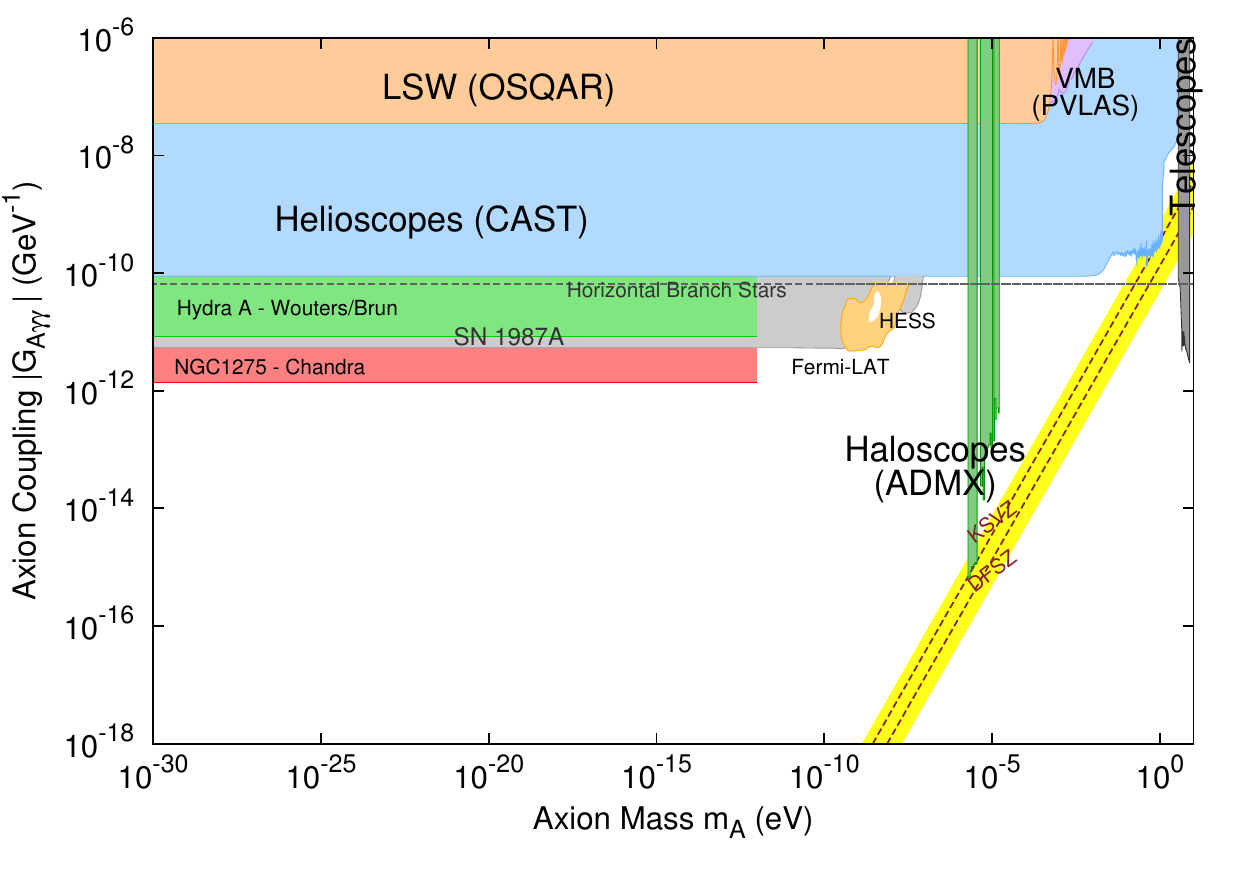}
\end{center}
\caption{For a central magnetic field strength $B_0=25\mu{\rm G}$, we show
the 95\% upper bounds on the ALP-photon coupling in comparison to previously obtained bounds on ALPs.}
\label{fig:exclusionplot}
\end{figure}

We scan over $g_{a \gamma \gamma}$ in steps of $10^{-13} \, {\rm GeV}^{-1}$. For the $g_{a \gamma \gamma}$ value excluded, we also check that the three values above it in our grid are also excluded. For the magnetic field parameters described above, we find $g_{a \gamma \gamma} \lesssim 1.4 \times 10^{-12} \, {\rm GeV}^{-1}$ which is shown in Figure~\ref{fig:exclusionplot}.

If we consider a more pessimistic scenario with $B_0 = 15 \, \mu {\rm G}$ and a minimum coherence length of $0.7 \, {\rm kpc}$, we instead find $g_{a \gamma \gamma} \lesssim 2.7 \times 10^{-12} \, {\rm GeV}^{-1}$. If we take an even more pessimistic scenario in which the central field is  $B_0 = 10 \, \mu {\rm G}$ and the minimum coherence length is $0.7 \, {\rm kpc}$, the bound increases further to $g_{a \gamma \gamma} \lesssim 4.0 \times 10^{-12} \, {\rm GeV}^{-1}$.

%%Finally, we note that if we could account for the features at 2.2 keV and 3.5 keV by an astrophysical or instrumental explanation, we expect that these bounds on ALPs would improve significantly (see Appendix~\ref{sec:signal}).

We finally note that if we excluded the feature at 2.2 keV from the fit, removing the 1.8 - 2.3 keV region based on this being the location of an effective area edge,
the bounds would improve significantly, to $g_{a\gamma\gamma} \lesssim 1.1 \times 10^{-12} \, {\rm GeV}^{-1}$ for the case of a central magnetic field of $B_0 = 25 \, \mu {\rm G}$
(see Appendix~\ref{sec:signal}). This illustrates the conservative nature of the bounds compared to any
mis-modelling of the actual data.

\subsection{Validation with MARX simulation}
\label{marx}

As a validation of these bounds, we also perform MARX simulations (\cite{MARX}) of the ACIS-I Edge observations to confirm that the results are robust against any pileup effects. Significantly piled-up data is known to produce complex effects, in particular with a bright background (see for instance~\cite{0004-637X-640-1-211}).

As the actual spectrum is a sum of the AGN powerlaw and thermal emission from the cluster, we first need to determine the magnitude of thermal
emission with the ACIS-I edge extraction region.

To extract the level of thermal emission within the ACIS-I Edge observations,
we follow two strategies. First, we took the longest on-axis ACIS-S
observation (\emph{Chandra} obsid 4952) and used the same extraction region as for the ACIS-I edge data (also excluding the AGN with
a central circle of radius 2 arcseconds). We fit an \texttt{apec} model to this, obtaining a temperature of $T \sim 4 {\rm keV}$,
an abundance of $Z \sim 0.4$ and a normalisation of 0.0022. While such a single-temperature \texttt{apec} model is not an accurate description of the gas dynamics at the centre of the cluster ($\chi^2/N = 474/375$), it is adequate for our purposes as we are more concerned with the level of pileup, and thus the overall \emph{magnitude} of thermal emission, rather than a fully perfect spectral shape.
Using this model together with the exposure and responses for the ACIS-I edge observations, this suggests the presence of around $\sim 30{\rm k}$ thermal counts present in the ACIS-I spectrum.

Secondly, as a check on this, we also used purely the ACIS-I edge data and directly fit it with a sum of an absorbed power-law and absorbed thermal emission. In doing this, we froze the temperature and abundance of the thermal model to the values obtained from the \emph{Hitomi} spectrum of the central region of the Perseus cluster, $T$ = 3.48 keV and $Z=0.54$ (\cite{Hitomi}).
This produced a best-fit normalisation for the \texttt{apec} norm of 0.0027, comparable to that attained when using the ACIS-S data to
determine the magnitude of thermal emission. This slightly higher value would corresponds to a total of $\sim 36$k counts
from thermal emission within the ACIS-I Edge spectrum. As both results are consistent, they show that of the $\sim 270$k counts in the ACIS-I edge spectrum only around 10-15\% arise from thermal cluster emission.

We then implement a MARX simulation of the ACIS-I edge observations. For this, we used the observational parameters (date, on-axis location and roll angle) for the longest ACIS-I edge observation of the AGN (\emph{Chandra} obsid 11713). As all the four actual ACIS-I Edge observations
have similar properties, we set the exposure of this MARX simulation to 200ks, the total exposure of the actual data.

We first simulate data without any axions present.
\begin{enumerate}
\item
The thermal emission was simulated as a spatially extended Gaussian with a width of 30 arcminutes, centred on the AGN.
It was modelled as an \texttt{apec} model with a temperature $T = 3.5 {\rm keV}$ and an abundance $Z=0.48$. The normalisation
of the model was adjusted so that it had the same fitted strength as for real data (as determined above)
when extracted within the region used for the ACIS-I Edge data.

\item
The AGN emission was simulated as a point source power-law. The normalisation of the power-law was adjusted so that -
after it had been combined with the thermal emission and the pileup processing applied - it had the same normalisation as the real data when extracted over the same extraction region.

\item
The two individual simulations were then combined using \texttt{marxcat} and the MARX pileup processing applied to the combined
simulation.

\item
Counts were grouped to 500 and the combined spectrum was fitted with the sum of a power law plus thermal emission. The previous two steps were
iterated until the fitted strength and index of the powerlaw matched that of the real data.
\end{enumerate}

We generated a total of 50 fake data samples in this way.
To match the fitted strength of the AGN ($4.7 \ti 10^{-3} \, {\rm ph} \, {\rm cm}^{-2} \, {\rm s}^{-1} \, {\rm keV}^{-1}$ at 1 keV)
required an intrinsic strength of $6.4 \ti 10^{-3} \, {\rm ph}\,  {\rm cm}^{-2} \, {\rm s}^{-1} \, {\rm keV}^{-1}$ at 1 keV.
Comparison of the number of counts within the extraction region in
the \texttt{marx} event files pre- and post-pileup processing suggests an overall pileup fraction of $15 \%$ in the dataset.
Nonetheless, the pileup is not sufficient to make the spectral fit unsatisfactory (the reduced $\chi^2$ values cluster around unity for $\sim 250$ degrees of freedom, with the largest reduced $\chi^2$ being 1.37).

As a test of an even more piled-up spectrum with a similar number of photon counts,
we also simulated data using a 100ks MARX exposure but with the intrinsic AGN strength
doubled. As this normalisation of $12.8 \ti 10^{-3} \, {\rm ph} \, {\rm cm}^{-2} \, {\rm s}^{-1} \, {\rm keV}^{-1}$ at 1 keV
is much larger than that present in the 2016 \emph{Hitomi} data, and as the AGN has been increasing in strength from
2001 onwards (see \cite{Fabian:2015kua}), pileup is much greater than in the actual 2009 data.
The fit now tends to be worse, but not terrible. Among the fifty fake data samples,
the average reduced $\chi^2$ is $\sim 1.25$ for $\sim 250$ degrees of freedom, but there are a few cases still with a reduced $\chi^2 < 1$.

We now repeat the MARX simulations including the effects of axions in the data. We multiply the AGN power law with a $P(\gamma \to \gamma)$
survival probability coming from photon-ALP conversion. As this leads to a net reduction in the number of photons present, we adjust the intrinsic
normalisation upwards so that the fitted normalisation matches the actual data.

For a fixed ALP coupling, we generate 50 fake data samples. To compare these results to those described above
using Sherpa's \texttt{fake\_pha} command, we simulate three separate ALP couplings: $g_{a\gamma\gamma}= 1, \, 1.5, \, 2 \ti 10^{-12} {\rm GeV}^{-1}$.
In all cases we assume the `optimistic' magnetic field model. We directly fitted the absorbed sum of a power law and \texttt{apec} thermal emission to the ACIS-I Edge extraction region for the fake data, and compared to the quality of the fit when doing so using the real data.\footnote{This is marginally different from the analysis procedure described in earlier sections, as it aims at fitting the thermal emission rather than subtracting it. We also did this analysis subtracting background emission from the real data set, and fitting only with a power-law.
The results however are very similar, with no significant changes to the excluded coupling.}
We first group counts to 500 and fit between 1 and 5 keV using the \texttt{chi2datavar} statistic.
Applied to the real data, this gives a reduced $\chi^2$ of 1.83 (coming from the excess around 2 keV).

We then apply this fit to data involving axions.
Of the fifty fake data sets produced using $g_{a\gamma\gamma} = 2 \ti 10^{-12} \, {\rm GeV}^{-1}$,
the lowest reduced $\chi^2$ was 4.07. When $g_{a\gamma \gamma} = 1.5 \ti 10^{-12} \, {\rm GeV}^{-1}$, the average
reduced $\chi^2$ is still $\sim 2.5$, but now with three out of fifty spectra with better fits than the
actual data $(\chi^2 / N < 1.83)$.
When $g_{a\gamma\gamma} = 1.0 \ti 10^{-12} {\rm GeV}^{-1}$, more than half the fake spectra have better fits than the actual data.
Using 95\% confidence level exclusions, this shows that $g_{a\gamma\gamma} = 2 \ti 10^{-12} {\rm GeV}^{-1}$ is strongly excluded
(the best scenario having a reduced $\chi^2$ of 4.07), while $g_{a\gamma\gamma} = 1.5 \ti 10^{-12} {\rm GeV}^{-1}$ is on the boundary of exclusion, and $g_{a\gamma\gamma} = 1 \ti 10^{-12} {\rm GeV}^{-1}$ is clearly not excluded.

This extremely sharp behaviour of the reduced $\chi^2$ with $g_{a\gamma\gamma}$ can be understood qualitatively.
As $P(\gamma \to a) \propto g_{a\gamma\gamma}^2$, a linear increase in $g_{a\gamma\gamma}$ leads to a quadratic increases in the amplitudes of modulations in the data.
For a Poissonian process, the amount of data required to detect a fixed fractional deviation grows quadratically with
the size of the deviation. So, roughly, a decrease in $g_{a\gamma\gamma}$ by a factor of two requires a sixteen-fold
increase in the quantity of data for the same statistical sensitivity.

We also repeated this using a 100ks exposure with an AGN that is twice as bright (to give a data sample that has a similar number of photons,
but is substantially more piled up than the actual data). It is again the case that $g_{a \gamma \gamma} = 2 \ti 10^{-12} {\rm GeV}^{-1}$ is strongly excluded, $g_{a \gamma \gamma} = 1 \ti 10^{-12} {\rm GeV}^{-1}$ is not excluded, and $g_{a \gamma \gamma} = 1.5 \ti 10^{-12} {\rm GeV}^{-1}$ is marginal.

We note that there are of course several differences between the MARX simulation and actual data processing.
The MARX treatement of pileup is simpler than the real physics of deposited electron clouds from the interaction of
photons with CCD chips. The MARX simulation also does not take into account Charge Transfer Inefficiencies present
on the actual chips, which obstruct the flow of charge to the readout.
However, these results imply that, in terms of bounding $g_{a\gamma\gamma}$, uncertainties due to pileup appear to be far smaller than uncertainties due to the cluster magnetic field.

\begin{figure}
\begin{center}
\includegraphics[width=0.85\textwidth]{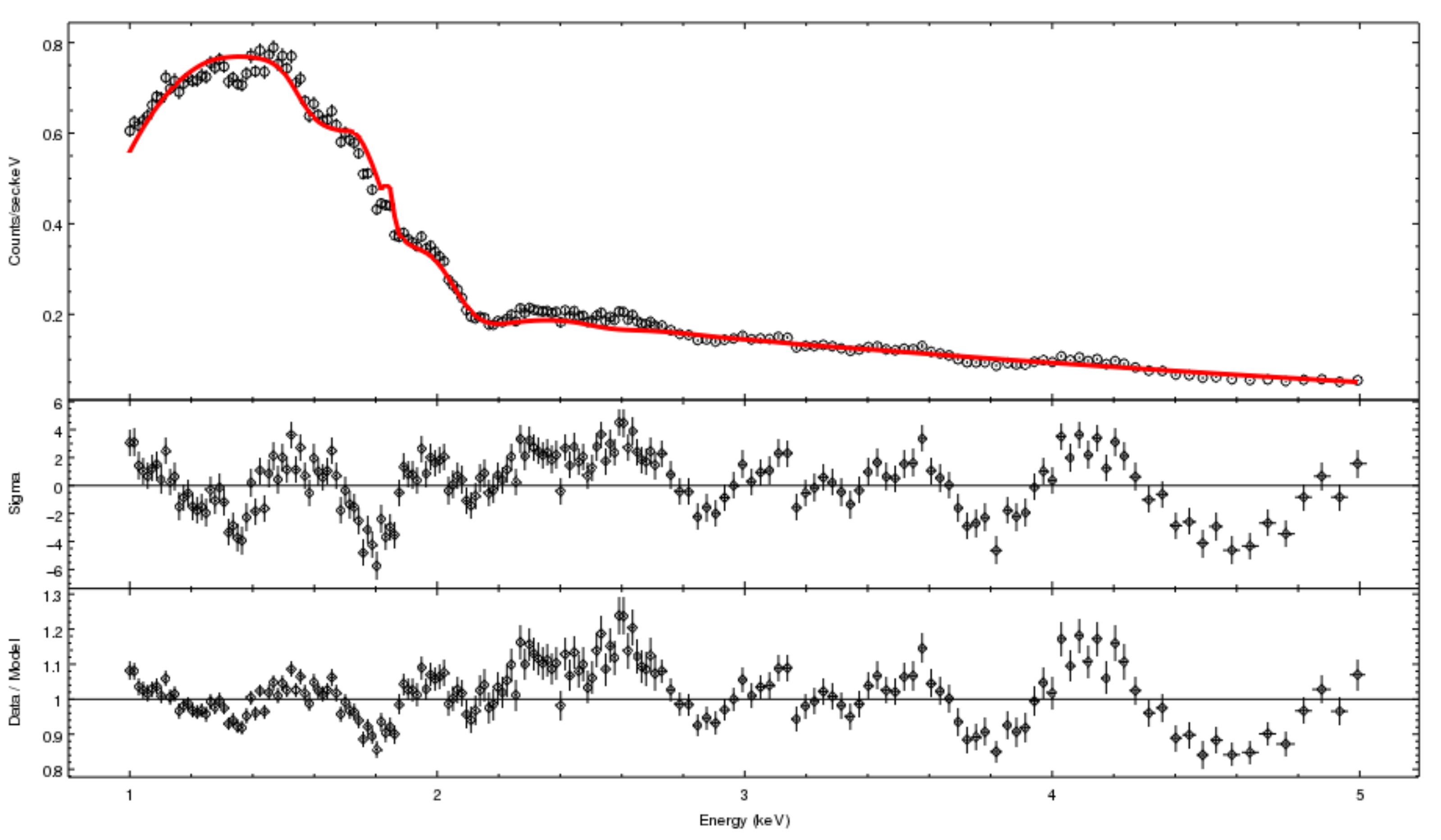}
\end{center}
\caption{An example of simulated data using \emph{ChaRT} and \emph{MARX} reproducing the observational conditions for the ACIS-I edge
observations, for an axion coupling of $g_{a\gamma\gamma} = 2 \ti 10^{-12} {\rm GeV}^{-1}$. In this case the spectrum is directly fit
with the sum of a power law and thermal \texttt{apec} emission (the ingredients to the simulation).
Despite the pileup fraction of $\sim 15 \%$,
the axion-induced modulations are clearly visible and cannot be fit by any conventional astrophysical model.}
\label{CHART}
\end{figure}

The above bounds used MARX's simulation of \emph{Chandra}'s optics.
The most advanced simulation of \emph{Chandra}'s optics is
through the \emph{ChaRT} simulator (\cite{CHART}), rather than MARX.
As \emph{ChaRT} requires the original source spectrum to be manually uploaded to a website, it is not possible to automate
this process to produce axion bounds (as every different ALP conversion template represents a different source spectrum).
However, for each of the couplings $g_{a\gamma\gamma}= 1, 1.5, 2 \ti 10^{-12} {\rm GeV}^{-1}$, we simulated some
individual \emph{ChaRT} spectra.
For the spectra simulated, the results are similar to those of the MARX simulations. We plot in figure
\ref{CHART} a sample spectrum for $g_{a\gamma\gamma} = 2 \ti 10^{-12} {\rm GeV}^{-1}$. The ALP induced modulations are clearly visible
leading to a large badness-of-fit.

\section{Conclusions}
\label{sec:conclusions}

The most basic point of this paper is that X-ray observations of NGC1275 are a superb way to search for ALPs. NGC1275 is an extremely bright -- and brightening --
X-ray point source shining through a galaxy cluster environment. If ALPs exist, they can lead to oscillatory modulations within the energy ranges
probed by \emph{Chandra} and \emph{XMM-Newton}.
Although there is uncertainty on the precise magnetic field structure along the line of sight to NGC1275, it is almost certainly the
case that current and future observations of NGC1275 provide a greater reach in
searches for light ALPs with $m_a \lesssim 10^{-12} {\rm eV}$ than even the proposed dedicated experiment IAXO~\cite{1401.3233}.

Previous \emph{Chandra} observations of NGC1275 already provide a dataset of extraordinary quality.
Three factors contribute to this. First,
 \emph{Chandra}'s angular resolution allows a large contrast between the AGN emission and cluster. Second, the existing observations are very deep, and include
 three independent locations of NGC1275 on the telescope. Third, it is a fortuitous fact that in one set of observations, NGC1275 was located on the edge of the chip, thereby
 providing a clean observation which minimises pile-up.

In this paper we have used all existing observations of NGC1275 with \emph{Chandra} and \emph{XMM-Newton} to search for spectral modulations induced by ALPs.
If they exist, ALPs give rise to spectral modulations, and we have used the absence of modulations at the $\mc{O}(30 \%)$ level to place leading current bounds on
ALP-photon couplings.

At the $\mc{O}(10 \%)$ level, there are
 two main modulations in the data -- one upward around 2--2.2~keV, and one downward around 3.4--3.5~keV.
 The former can be attributed to the Iridium edge in the mirror.
In the main body of the paper, we have focused on constraints that can be placed on ALP parameter space. In Appendix~\ref{sec:signal}, we discuss
residuals in greater detail. In particular, the residual around 3.5 keV can be well described as an absorption feature at $(3.54 \pm 0.02)$ keV, the same energy as the diffuse
cluster excess observed in \cite{Bulbul} and \cite{Boyarsky}.

The existing dataset already places powerful constraints on ALPs.
Nonetheless, from the perspective of ALP physics the dataset could be substantially improved with even relatively modest further observations.
NGC1275 is now brighter than it was in 2009 by a factor of two. Observations with either \emph{XMM-Newton} in Small Window mode or \emph{Chandra} off-axis and in reduced frame time
could give both a larger and cleaner dataset than the best used in this paper, the 2009 ACIS-I edge observations.
X-ray observations of the centre of the Perseus cluster therefore provide an outstanding way to look for new physics beyond the Standard Model.

\section*{Acknowledgments}

%%CHANGE JC
We thank David Marsh and the referee for comments on the paper, Alexis Finoguenov and Jeroen Franse for helpful discussions, and the Bethe Centre for Theoretical Physics for hospitality while part of this work was carried out. We would like to thank Gray Rybka for providing his exclusion plot template. We also thank the staff and science teams of the \emph{Chandra} and \emph{XMM-Newton} X-ray observatories for documentation
that is sufficiently clear that particle theorists can follow it and friendly helpdesks. This project is funded in part by the European Research Council starting grant `Supersymmetry Breaking in String Theory' (307605). JC is also funded by a Royal Society University Research Fellowship.
\software{CIAO v4.7~\cite{ciao}, Sherpa~\cite{sherpa}, HEASOFT v6.17, SAS v15.0.0, MARX 5.3.2, ChaRT v2.}

\appendix

\section{Modelling Pile-up}
\label{subsec:Pileup}

The main text addressed pile-up by removing central, high-pile-up regions from the extraction region. Here we take a complementary approach, aiming to
model the pile-up directly. While this modelling does not provide a complete picture of pile-up, it ameliorates its effect. A full treatment
would require customised tools going beyond the scope of this paper (the pile-up tool \texttt{jdpileup}~\cite{Davis:2001} provided with Sherpa is optimised for on-axis sources, while NGC1275 is off-axis in all ACIS-I observations).

We first describe our pile-up modelling for the ACIS-S observations, where NGC1275 is on-axis. This results in a high degree of pile-up -- the central pixels have large
numbers of counts above 10 keV, all of which arise from multi-photon pile-up events.

We use the \texttt{jdpileup} model as described in~\cite{Davis:2001}.
As this model assumes the source is on-axis, it is directly appropriate here.
The model assumes Poisson statistics to calculate the probability of different numbers of photons hitting an event-detection cell (a 3$\times$3 pixel region in Chandra) within the read-out time adjusted to these respective observations. For the ACIS-S and ACIS-I midway observations the read-out time was 3.1 seconds, and for the ACIS-I edge observations it was 3.2 seconds.
The model then convolves this probability with the probability of such events being assigned a `good' grade, and the conversion from photon energy to pulse height by the detector. Obviously this is a non-linear process: the pile-up in any particular bin depends on the energy spectrum for all energies below the bin. The parameters of the model therefore need to be determined together with those for the spectrum under consideration. This can lead to degeneracies in parameter space, in particular for a simple power law.

As per \cite{Davis:2001} the two parameters of the pile-up model that we allow to vary are $\alpha$ and $f$, where $\alpha^{p-1}$ is the probability that $p$ piled photons will be assigned a `good' grade, and $f$ is the fraction of events to which pile-up will be applied. The other parameter that will prove important is $n$, the number of regions to which \texttt{jdpileup} will be applied independently. For a point source this should be set to 1, as was done for the ACIS-S observations. For an extended source it should roughly correspond to the number of 3$\times$3 pixel islands in the region. The reason for this is that \texttt{jdpileup} assumes spatial uniformity across the extraction region. For the ACIS-I midway observations, where the AGN is smeared out across several pixel islands, the value of $n$ proved difficult to determine for large extraction regions, with the value corresponding to the best fit being unrelated to the number of pixel islands. We therefore constrained ourselves to a smaller
central
region,
with little variation in count rate between pixels, to give us more control over the pile-up model.

It is also worth noting that the \texttt{jdpileup} model is set to zero for energies less than 0.5~keV and energies greater than 10~keV.
As there are many events above the 10~keV range for the ACIS-S observations, we only model the spectrum up to 10~keV, as extending the fit beyond that
would result in the model parameters being sent to unphysical values.

For the ACIS-S pile-up analysis, we used a circular region around NGC1275 of radius 3.5 arcseconds, with the
background taken from an annular region of inner radius 4 arcseconds and outer radius 7 arcseconds.
Counts were grouped to 1500 per bin.
The spectrum was modelled using an absorbed power law and thermal emission with temperature $T=0.85 {\rm keV}$, and fitted with the \texttt{jdpileup} model for energy values between 1 and 10~keV.
The data was fit using the \texttt{moncar} Monte-Carlo method, and run several times to ensure the global minimum had been found.

\begin{figure}
\begin{center}
\includegraphics[width=0.8\textwidth]{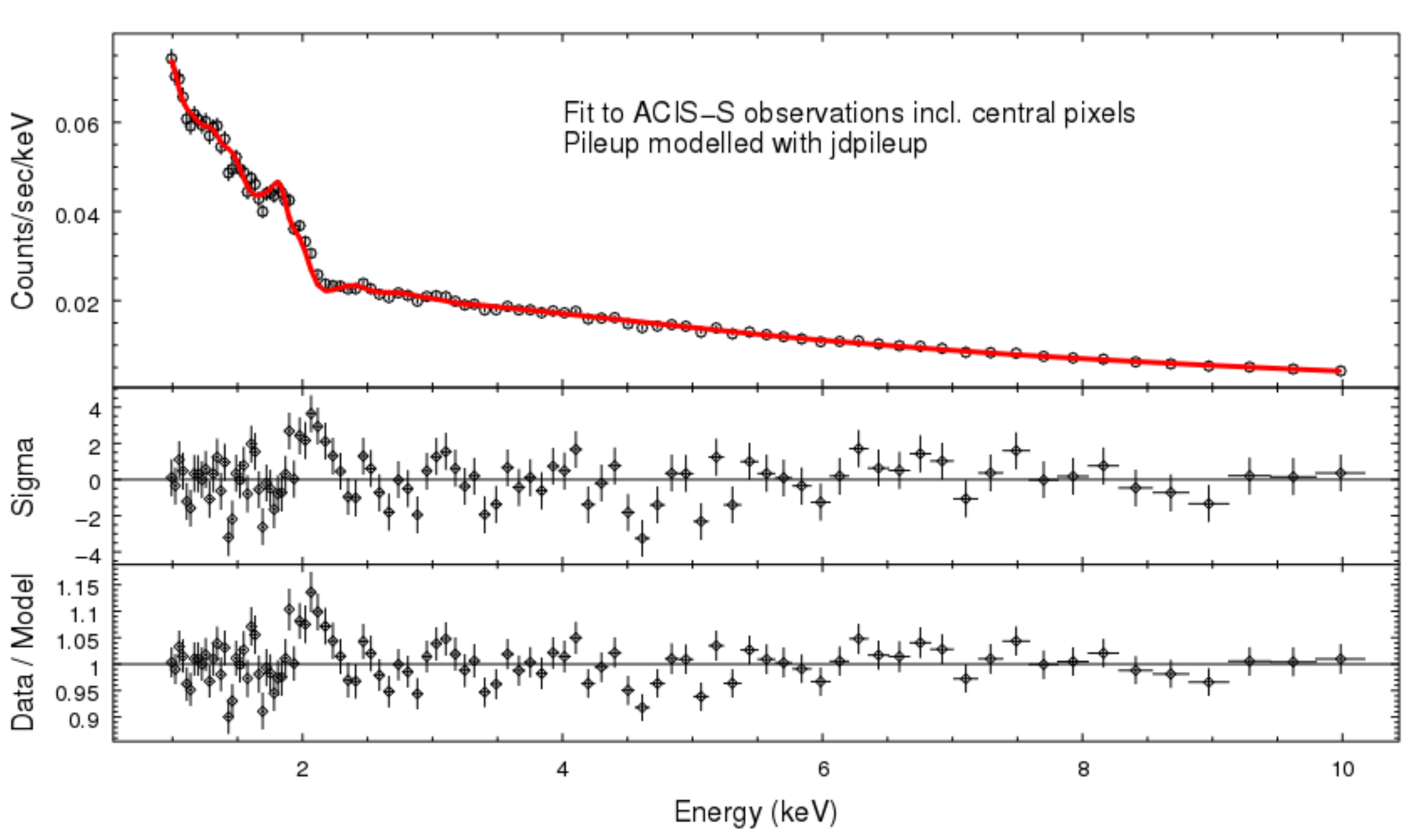}
\end{center}
\caption{The ACIS-S observations, involving 177000 counts after background subtraction. The fit is to an absorbed power law with a thermal component of $T=0.85$ keV, and pile-up is modelled with \texttt{jdpileup}. The ratio of data to model is shown at the bottom of this figure.}
\label{fig:pileup_s}
\end{figure}

The resulting fit and the ratio of data to model are shown in Figure~\ref{fig:pileup_s}. The best fit parameters involve
$n_H=2.6 \ti 10^{21} {\rm cm}^{-2}$ and a spectral index $\gamma = 1.81$.\footnote{As the uncertainties on the pile-up modelling
are hard to quantify, we only quote best-fit parameters and do not include errors.}
The best-fit $\alpha$ and $f$ parameters of \texttt{jdpileup} were $\alpha = 0.660$ and
$f = 0.943$ respectively. We can clearly see that this model provides a reasonable description of the data all the way up to 10~keV, and produces physically sensible values for $n_H$ and $\gamma$, despite the model estimating a pile-up fraction of 82\%. The reduced statistic of the fit is 1.75, with a Q-value of $10^{-5}$.
%As before, we again see a clear excess around 2.0--2.2~keV, and a low-significance dip at 3.4--3.5~keV.
%There is also a narrow local deficit around 4.6~keV at around 3-sigma significance. The 6.3~keV iron line (6.4~keV in the cluster rest frame) is obscured by the high level of pile-up in that energy region (see also Appendix~\ref{sec:appb1} for more details on the iron line).
While we re-emphasise that this will not represent a perfect account of pile-up, it does capture the relevant physics, producing a sensible fit with physical parameters.

For the ACIS-I midway observations we considered a central 6$\times$6 pixel extraction region (this is almost exactly the complement of the clean ACIS-I midway spectrum used in the previous subsection). The resulting spectrum contains around 136000 counts, reducing to 134000 after background subtraction, giving a very high AGN to cluster contrast of 67:1. After background subtraction, the fraction of counts in the 7-10~keV band and 10-15~keV bands are 5.3\% and 1.2\% respectively. Counts were grouped to 1000 per bin. The spectrum was modelled using an absorbed power law and fitted with the \texttt{jdpileup} model, this time for energy values between 1 and 9~keV, to ensure no counts with energy greater than 10~keV were included in the final bin. The \texttt{jdpileup} parameter $n$ was set to 4, the number of 3$\times$3 pixel islands in the extraction region.

\begin{figure}
\begin{center}
\includegraphics[width=0.8\textwidth]{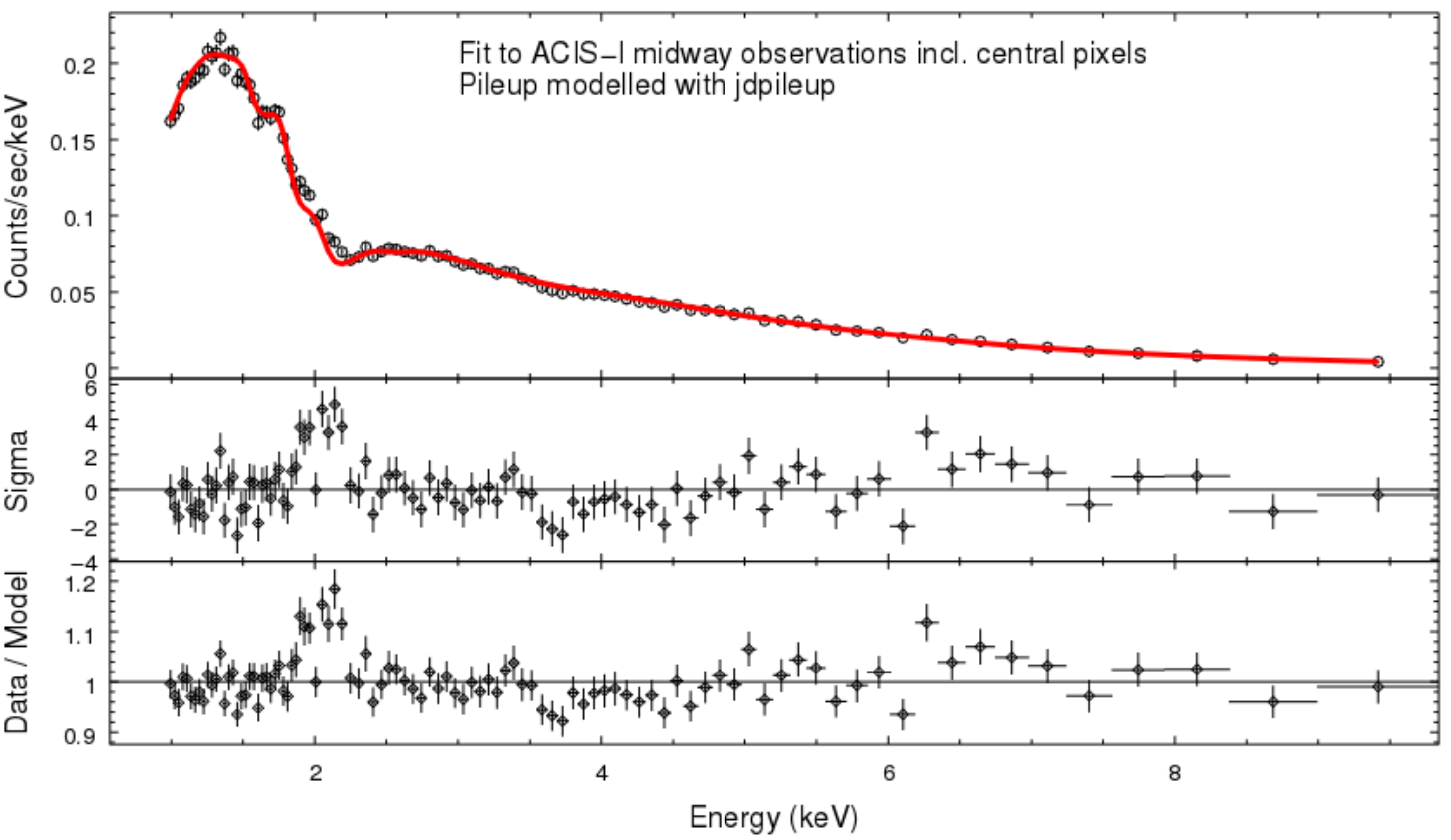}
\end{center}
\caption{The ACIS-I midway observations from a central 6$\times$6 pixel region, involving 134000 counts after background subtraction. The fit is to an absorbed power law and pile-up is modelled with \texttt{jdpileup}.}
\label{fig:pileup_mid}
\end{figure}

The resulting fit and ratio of data to model are show in Figure \ref{fig:pileup_mid}. The best fit parameters involve $n_H=2.5 \ti 10^{21} {\rm cm}^{-2}$ and a spectral index $\gamma = 1.93$. The best-fit $\alpha$ and $f$ parameters of \texttt{jdpileup} were $\alpha = 0.324$ and
$f = 0.975$ respectively, and the estimated pile-up fraction was 35\%. While the fit is not perfect, it does give a reasonable characterisation of the data.%Once again the excess around 2.0--2.2~keV is apparent, along with a small deficit at 3.4--3.5~keV.
%The (unmodelled) 6.3~keV iron line is also visible as an excess in the data. Again, the fit it good up to 9~keV.

We finally consider the case of the ACIS-I edge observations, where pile-up is relatively weak.
Although a reasonable fit can be made with no pile-up modelling,
we now aim at modelling the pile-up also in these observations
for completeness.
There is a rule of thumb\footnote{{\tt cxc.harvard.edu/csc/memos/files/Davis\_pileup.pdf}}
that 0.007 counts/second per 3$\times$3 pixel island is  about 1\% pile-up,
and 0.07 counts/second is about 10\%.
We will assume that the ACIS-I edge observations
with mild cleaning have pile-up fractions of 10\% or lower.

For the ACIS-I edge observations,
we used as source an elliptical annulus of radii 3 and 5 arcseconds,
 with the 12 brightest pixels removed by a contour.
The background was extracted from an elliptical annulus of outer
radii 18 and 23 arcseconds, with
inner radii 8 and 12 arcseconds.
The resulting spectrum contains around 128000 counts, reducing to 111000 after background subtraction, giving an AGN to cluster ratio of 6:1. After background subtraction, the fraction of counts in the 7--10~keV band and 10--15~keV bands are 1.1\% and 0.1\% respectively.
We grouped counts to 1000 per bin.
Here there are fewer counts out at higher energies than in the previous two sets
of observations,
so we fit only out to 5~keV.
 The \texttt{jdpileup} parameter $n$ was set to 16, the number of 3$\times$3 pixel islands in the extraction region.
This should only be thought of as a rough estimate, as the \texttt{jdpileup} model is stated
to be accurate for on-axis point sources, and
  the edge observations are also not spatially uniform in terms of counts.
 The fit is to an absorbed power law, and results in $n_H = 2.7 \ti 10^{21}{\rm cm}^{-2}$ and a spectral index
of $\gamma=1.89$.
The best-fit $\alpha$ and $f$ parameters of \texttt{jdpileup} were $\alpha = 0.52$ and
$f = 0.87$ respectively. The pile-up fraction was 3\%, consistent with estimates
and very mild as expected. %The excess around 2.0--2.2~keV is again apparent, as well as the small deficit at 3.4--3.5~keV.
\begin{figure}
\begin{center}
\includegraphics[width=0.8\textwidth]{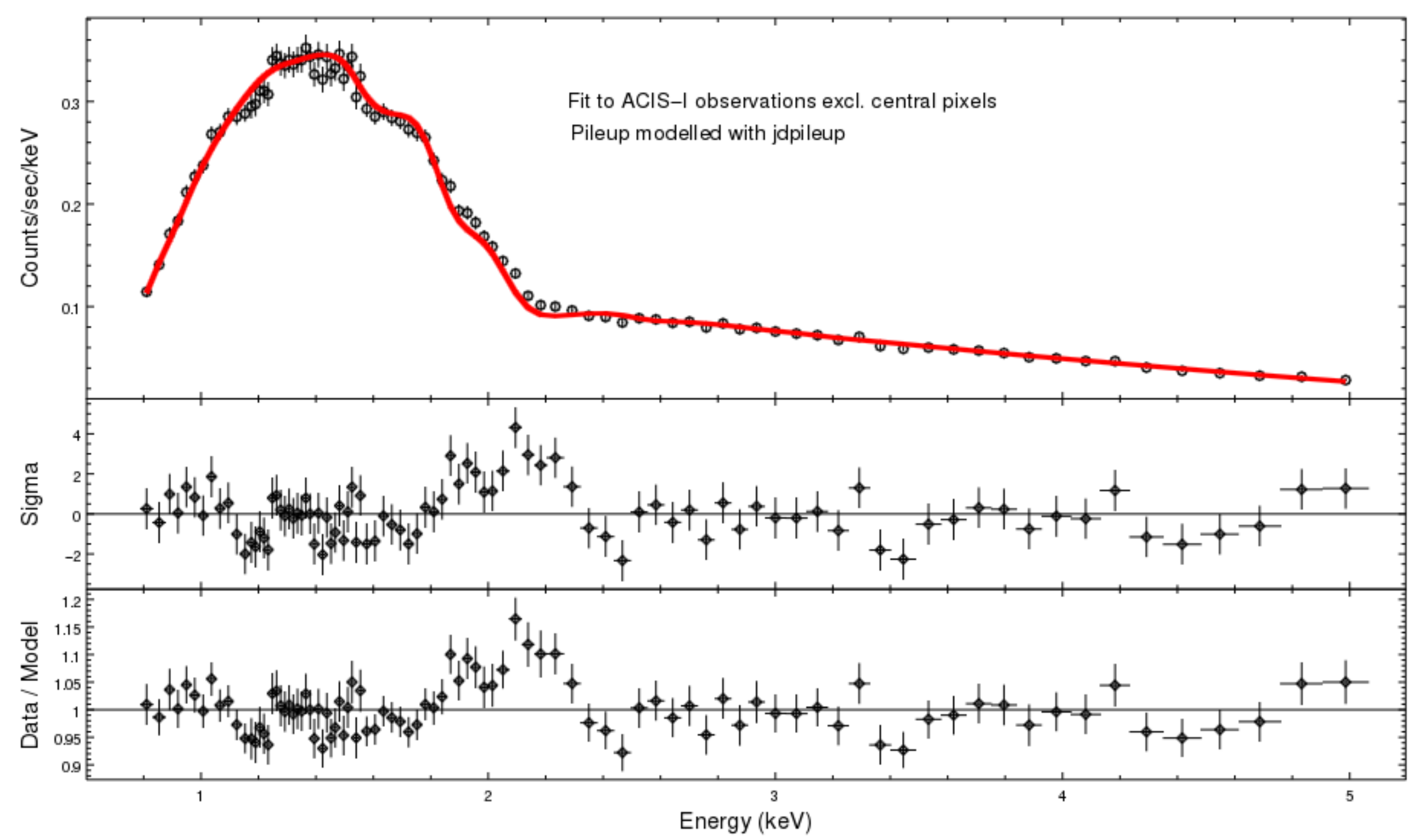}
\end{center}
\caption{The ACIS-I edge observations, involving 110000 counts after background subtraction.
The fit is to an absorbed power law and pile-up is modelled with \texttt{jdpileup}. }
\label{fig:pileup}
\end{figure}

\section{Analysis of $10\%$ residuals}
\label{sec:signal}
No residuals in the spectra exceed 10\%. At the 10\% level,
the \emph{Chandra} spectra presented in the main body of the paper show
two main features departing from a power law: an excess at 2--2.2~keV, present
in all \emph{Chandra} observations with overwhelming statistical significance, and a deficit 3.4--3.5~keV that is not as strong but is still present at almost five (local) sigma
in the ACIS-I edge data. These features ensure that an absorbed power law (plus thermal component) is not a good fit to the data.

First, we perform a statistical analysis of these features at $E \sim 2 {\rm keV}$ and $E \sim 3.5 {\rm keV}.$ In the second part we mention potential instrumental, astrophysical or new physics explanations.

%Given that the nature of ALP physics is to produce localised modulations in the data, we cannot simply dismiss them and are obliged to consider them as possible signals.
%We start by considering their purely statistical significance.

\subsection{Statistical Significance of Features at 2 and 3.5 keV}
\label{sec:appb2}

We first consider the $E \sim 2 {\rm keV}$ feature.
Considering the clean ACIS-I Edge data, we include a positive Gaussian $(xswabs * (xspowerlw + xsgaussian))$ and analyse its effect on the fit.
For a zero-width Gaussian, this leads to an improvement in the fit by $\Delta \chi^2 = 33.7$ with a best-fit energy of 2.13 keV.
For a finite-width Gaussian, the improvement is $\Delta \chi^2 = 48.0$, with a best-fit energy of 2.02 keV and a best-fit width of 0.2 keV.

For the ACIS-I midway data, a similar additional Gaussian improves the fit by $\Delta \chi^2 = 18.6$ with a best-fit energy of 2.17 keV (the fitted width is
much smaller than the detector resolution and so results are identical for zero-width and finite-width Gaussians). For the ACIS-S data, the additional Gaussian
improves the fit by $\Delta \chi^2 = 20.3$ with a best-fit energy of 2.06 keV (again, fitting the width gives a result smaller than the detector resolution and so does
not affect the result).

%These features are much larger than any other residuals in the spectrum (there are no other significant positive residuals), and they confirm the \emph{statistical} significance of this excess (as this excess occurs at the location of an effective area dip, it is important to consider systematic effects).

\begin{figure}
\center
\includegraphics[width=0.9\textwidth]{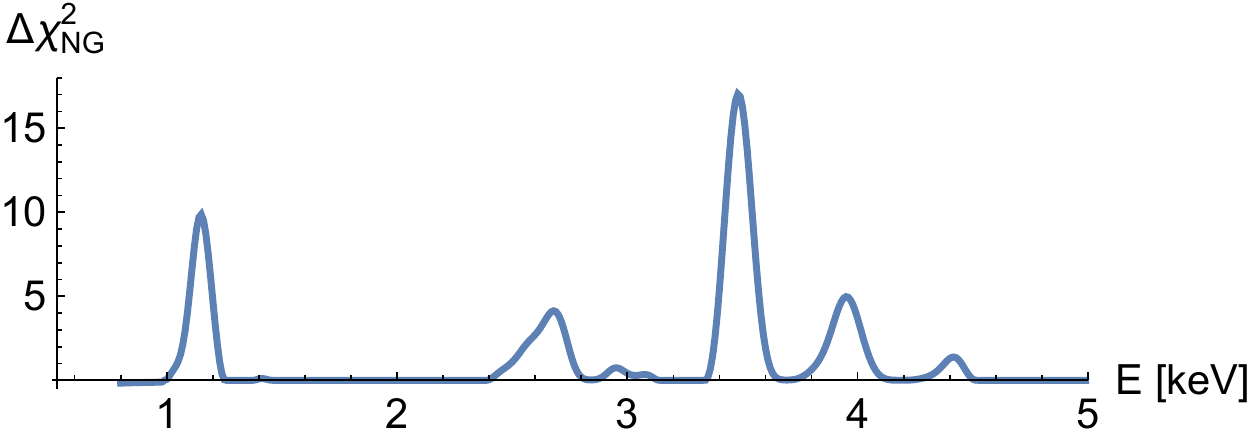}
\caption{The improvement in $\chi^2$ attainable by adding a negative Gaussian at the specific energy for ACIS-I Edge observations.} \label{RadialProfile1}
\end{figure}

\begin{figure}
\center
\includegraphics[width=0.9\textwidth]{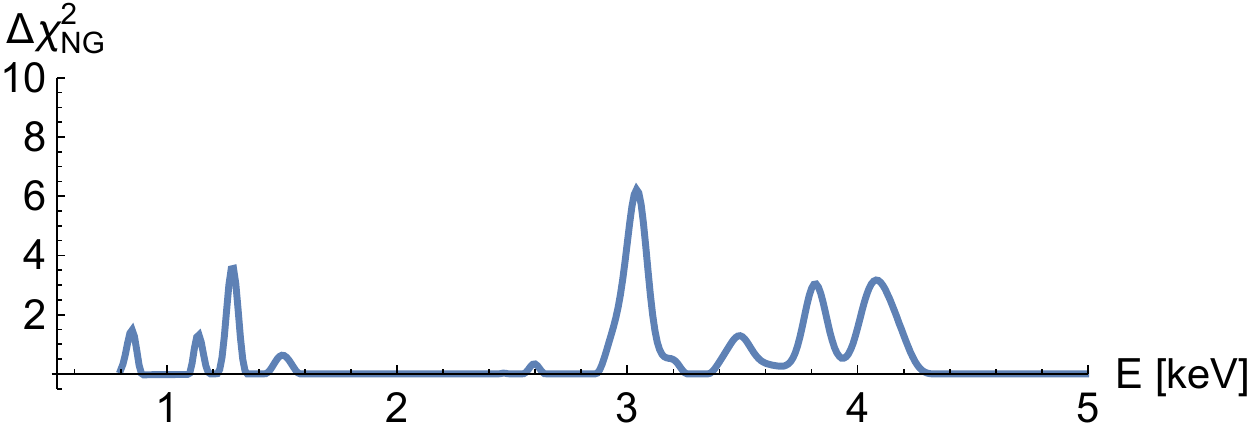}
\caption{The improvement in $\chi^2$ attainable by adding a negative Gaussian at the specific energy for ACIS-I Midway observations.} \label{RadialProfile2}
\end{figure}

\begin{figure}
\center
\includegraphics[width=0.9\textwidth]{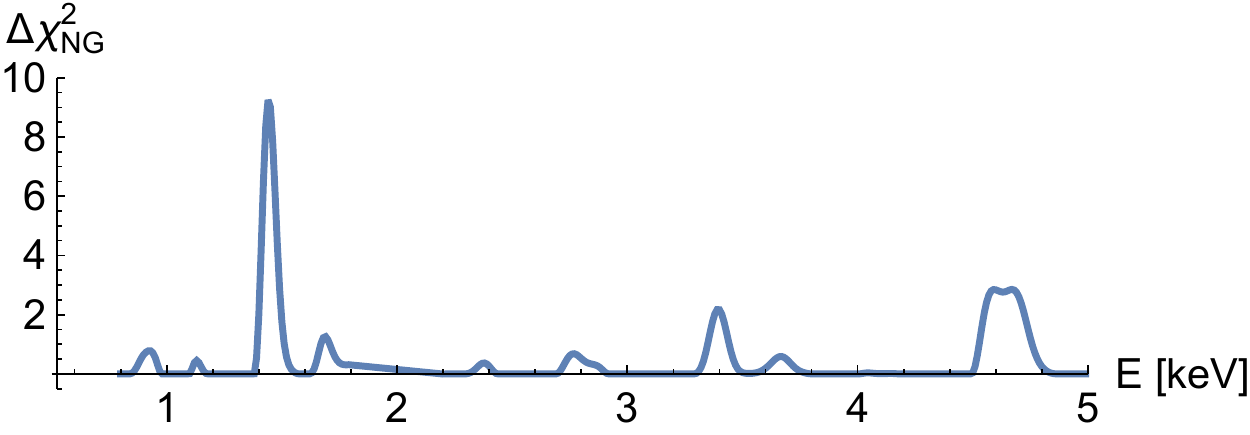}
\caption{The improvement in $\chi^2$ attainable by adding a negative Gaussian at the specific energy for ACIS-S observations.} \label{RadialProfile3}
\end{figure}

For the 3.5 keV feature, we consider the cleaned datasets for the three sets of observations and include a negative Gaussian $(xswabs * (xspowerlw - xsgaussian))$. We formally treat the Gaussian as zero-width, but any finite width much narrower than the ACIS energy resolution gives an identical result. For the three data sets (ACIS-I Edge, ACIS-I Midway, and
ACIS-S) we plot below (Figures~\ref{RadialProfile1}-\ref{RadialProfile3}) the improvement in the $\chi^2$ that can come from adding an additional negative Gaussian $\Delta\chi^2_{\rm NG}.$ The ACIS-I Edge data show a strong preference for an additional negative Gaussian ($\Delta\chi^2\sim17$) at 3.5 keV. For both the ACIS-I Midway and ACIS-S data, a negative Gaussian at 3.5 keV mildly improves the fit ($\Delta \chi^2 \sim 1.5$ and $\Delta \chi^2 \sim 0.4$) but is not required (as the ACIS-I Edge dataset is both larger and cleaner than the other two datasets, these results are consistent -- in particular, the inferred \emph{strength} of the dip is consistent within 2 $\sigma$).%%CHANGE

Across all three plots, the largest feature is clearly seen to be the residual in the ACIS-I Edge data at 3.5 keV.
The next strongest residuals are two features at $E \sim 1.2~{\rm keV}$ (ACIS-I Edge) and
$E \sim 1.4~{\rm keV}$ (ACIS-S) with $\Delta \chi^2 \sim 10$. In both cases, there are very strong atomic lines visible in the
background spectra at precisely these energies (Fe XXI, Fe XXII, Fe XXIII, Fe XXIV, Mg XII). These features can be reliably associated with these atomic lines, coming from a small mis-subtraction of the background (deep in the core of the Perseus cluster, the physical conditions of the gas in the signal region and background region will not be precisely identical).

\begin{figure}
\center
\includegraphics[width=0.6\textwidth]{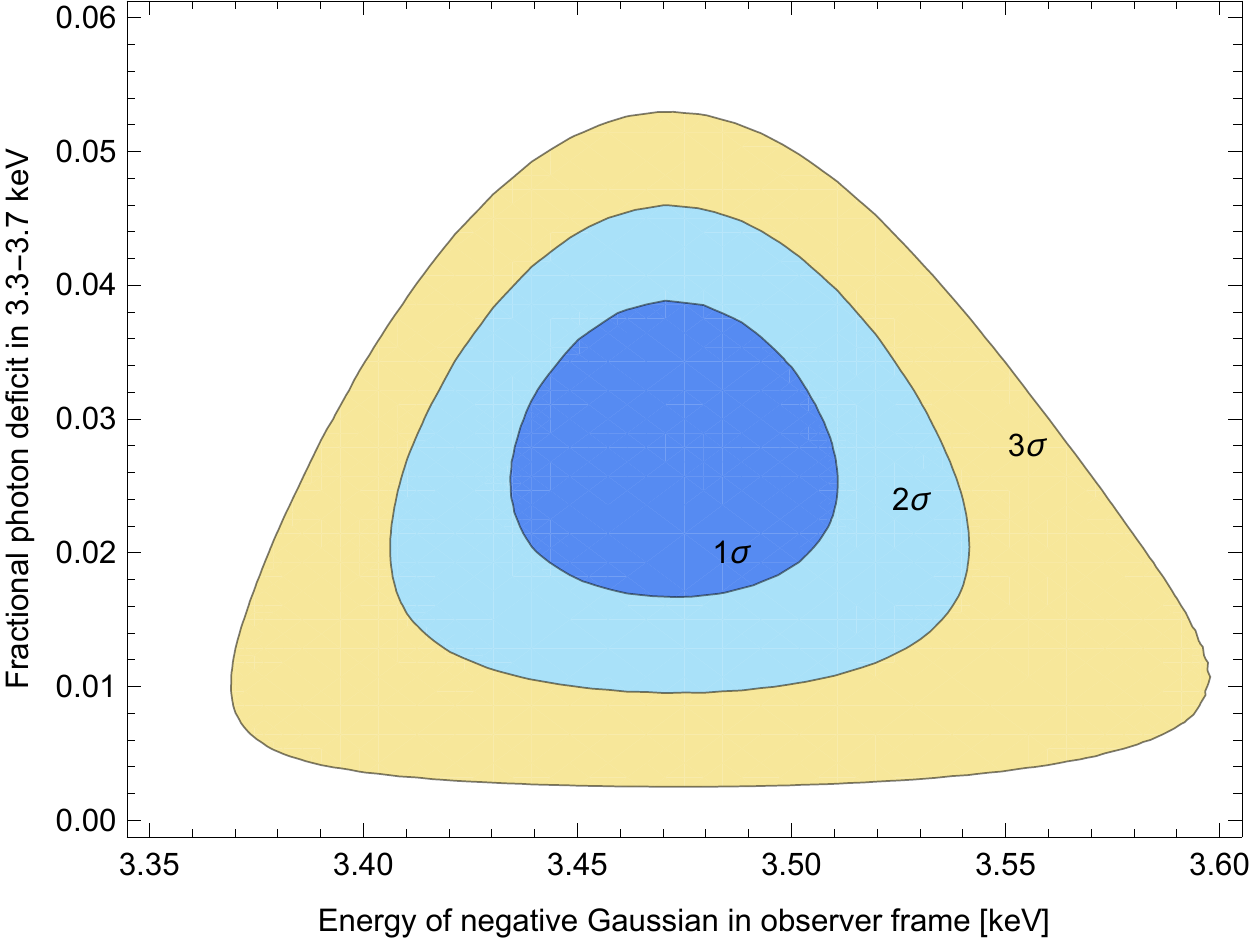}
\caption{The overall significance and location for the 3.5 keV deficit, combined across all \emph{Chandra} observations (cleaned datasets only). We have used the induced
fractional deficit in photons in the 3.3 - 3.7 keV region to facilitate a uniform comparison
for the strength of the dip, as the AGN luminosity varies between observations.} \label{RadialProfile}
\end{figure}

In Figure~\ref{RadialProfile}, we produce a combined plot of the significance of the 3.5 keV deficit across all \emph{Chandra} observations.
In order to compare observations taken at different times and with different intrinsic AGN luminosities, to measure the strength of the deficit
we have used the induced fractional deficit of photons in the 3.3 -- 3.7 keV regime. The feature remains over 3$\sigma$ significant globally, and is also located at
precisely the same energy as the diffuse cluster excess observed in \cite{Bulbul, Boyarsky}.

%%\subsection{Possible explanations for residuals}

\subsubsection{The 2.2 keV residual}

Despite the overwhelming statistical significance of the feature at 2-2.2keV, it is at the same location of
a sharp effective area edge from the Iridium coating of the mirrors.
As pile-up arising from high flux levels can generate fake excesses at the location of such edges,
we associate the feature with this edge and do not discuss it further.

\subsubsection{The 3.5 keV deficit}
We now consider possible explanations for the feature at 3.4--3.5~keV, present as a deficit of data compared to model.
This is present at very high significance in the ACIS-I edge observations. Given the smaller data samples present there,
the ACIS-I midway and ACIS-S observations are compatible but are not significant in themselves.

While this feature lacks the statistical significance of the one at 2--2.2~keV, it is harder to
come up with instrumental or astrophysical explanations for such a deficit.
Compared to the 2--2.2~keV region, at 3.4--3.5~keV the \emph{Chandra} effective areas are
smooth functions of energy, and in general this is a clean part of the spectrum, so the previous arguments for pile-up contamination or effective area miscalibration do not apply here.

The simplest explanation is as a statistical fluctuation -- although as this deficit is present as a local $\sim 4.5 \sigma$ deficit in the ACIS-I edge observation,
such an explanation is problematic. It is also notable that this occurs at precisely the same energy as for the diffuse excess found in
\cite{Bulbul, Boyarsky}.

In terms of conventional astrophysical explanations, the most obvious candidate for this deficit would be an atomic
absorption line. However it is difficult to see
how this could work, as there are no strong lines around this energy. Even more seriously, absorption along the sightline to NGC1275
dominantly arises from the Milky Way -- this is evidenced by the fitted values of $n_H$, which are all consistent with the galactic
value towards Perseus of $n_H \simeq 1.5 \ti 10^{21} {\rm cm}^{-2}$. This implies that, if the feature at 3.4--3.5~keV did arise from
atomic absorption, a similar feature should also arise for the continuum cluster spectrum of Perseus, as
the galactic $n_H$ values have degree-scale gradients and so $n_H$ is approximately the same for all sightlines to Perseus.
However, no such absorption feature at 3.5~keV is detected in the cluster spectra of Perseus --- indeed, precisely the opposite is found and instead a significant excess
is found at 3.5~keV~\cite{Bulbul, Boyarsky} (for a review see~\cite{1510.00358}).

The most plausible conventional explanation of the 3.4--3.5~keV deficit is then as a localised
detector effect or statistical fluctuation for the ACIS-I edge observations, combined with mild downwards statistical fluctuations for the ACIS-S and ACIS-I central observations.
However, we also note that inspection of the background region for the ACIS-I edge observations does not show any deficit around 3.4--3.5~keV.

\subsubsection*{New physics explanations}
Given the observations of~\cite{Bulbul, Boyarsky} of an unidentified line at 3.5 keV in diffuse emission from the Perseus cluster,
it is an interesting fact that we observe a deficit at $\sim$ 3.5~keV in our observations of the NGC1275 AGN at the centre
of the cluster. This of course may be just a coincidence, but
in the context of new physics models we mention two ways that these facts could be more than simply a coincidence.

The first involves models of excited dark matter invoked for the 3.5~keV
line, where dark matter has a resonance at an energy $\sim$ 3.5~keV above its ground
state (for example as in~\cite{Finkbeiner, 14080233, Cline, 1501.03496, 1503.03057, Profumo}).
In this case there is then an absorption cross section of $E \sim 3.5 {\rm keV}$ photons on dark matter.
While for an isotropic initial distribution of photons such absorption and re-emission
would not affect the photon spectrum, for a directional beam dark matter absorption will result in an absorption hole in the spectrum (the presence of
an absorbing torus around an AGN ensures its outward radiation is indeed directional).

Would such an effect have been observed already elsewhere? We do not see why. In this scenario,
the relevant quantity determining the fractional absorption rate
is the dark matter column density along the line of sight. It is entirely plausible that the dark matter column density towards
 the NGC1275 AGN is larger than for almost any other direction in the universe. This is because the emission all originates very close to the central AGN, and so the column
density is sensitive to not just the Perseus cluster, but also the central cluster galaxy NGC1275 right down to any sub-pc level dark matter spikes close to the central
supermassive black hole. The effect we observe is not large -- a 10\% reduction over around 100 eV in width -- and requires a spectrum with
$\mc{O}(10^5)$ counts for a statistically significant detection. With a smaller dark matter column density, the effect would reduce to an unobservable
$\mc{O}(1 \%)$ effect.

There is a second possible connection to the 3.5~keV line.
An attractive scenario for the 3.5~keV line involves decay of dark matter to 3.5~keV ALPs, which then convert to photons through axion-photon conversion
in the cluster magnetic field~\cite{14032370, 14047741, 14065518, 14101867}.
In this scenario the strength of the 3.5~keV line depends on the efficiency of ALP-photon interconversion -- and so broadly is
expected to be larger in regions with large magnetic fields extended over wide areas: for example, in galaxy clusters as opposed to galaxies.
The 3.5~keV line is observed to be stronger towards the centre of Perseus than for other clusters; one way this could arise is
if it fortuitously happens that
ALP-photon interconversion is particularly efficient around 3.5~keV for sightlines towards the centre of Perseus.
In this case, the presence of a deficit of $E \sim 3.5 {\rm keV}$ photons from NGC1275 and an excess of $E \sim 3.5 {\rm keV}$ photons from
the cluster as a whole could come from the same underlying physics -- efficient
photon-ALP interconversion at energies $E \sim 3.5 {\rm keV}$ along sightlines towards the centre of Perseus.

\subsubsection{ALP Interpretation of Features}

The purpose of this paper was to use the extraordinary dataset of counts from the NGC1275 AGN to search for spectral irregularities, with the intent
of constraining ALP parameters. This search has resulted in two features being present in the data at high statistical significance,
the former of which is consistent with arising from pile-up around the Iridium edge.

\begin{figure}\begin{center}
\includegraphics[width=0.8\textwidth]{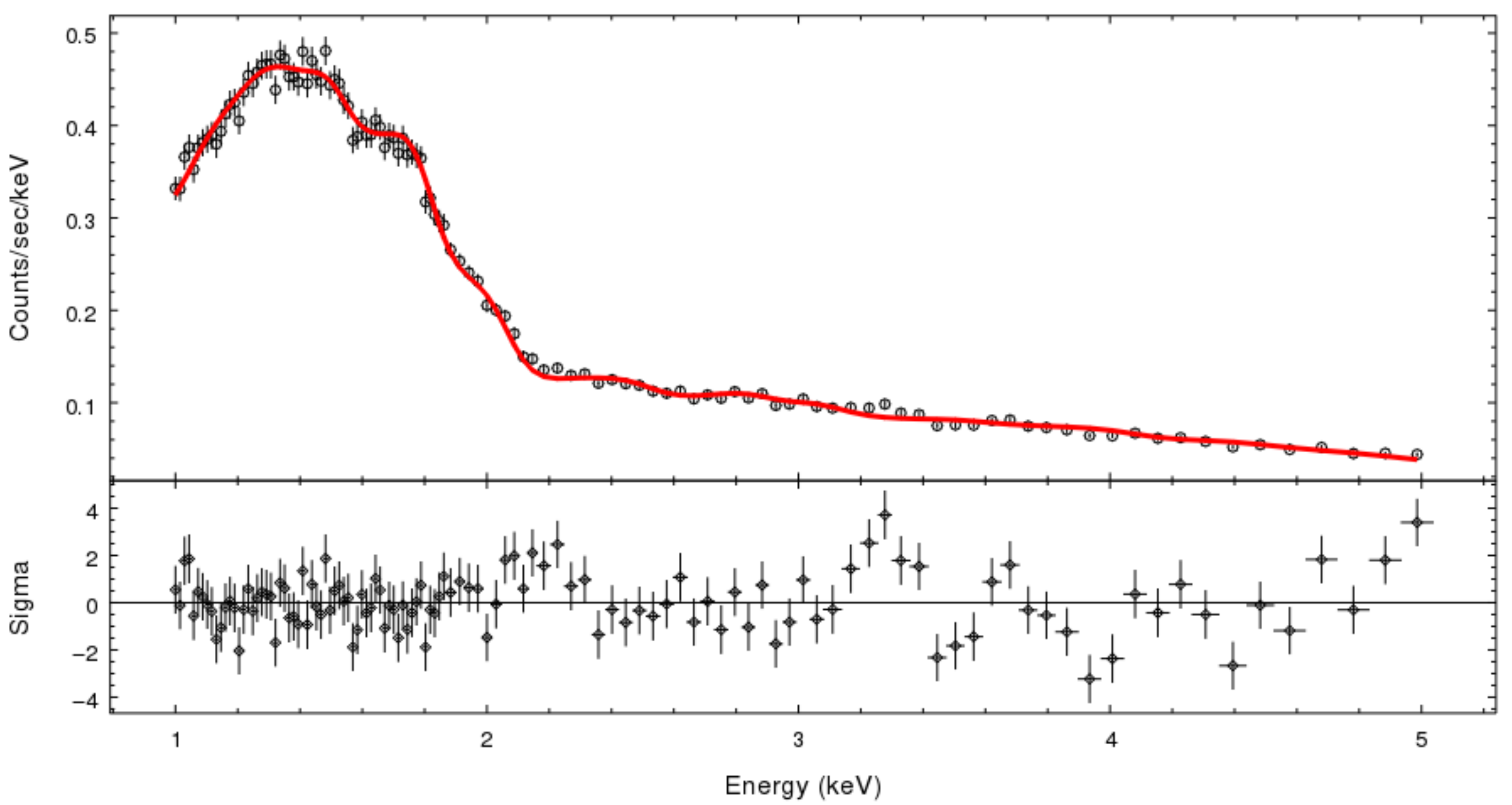}
\end{center}
\caption{A fit to the clean ACIS-I edge observations with an absorbed power law multiplied by the photon survival probability for a
specific example of ALP-photon conversion. We include %%JC
ALPs with $g_{a \gamma \gamma} = 1 \times 10^{-12} \, {\rm GeV}^{-1}$ and assume a central field of $B_0 = 25 \, \mu {\rm G}$. The reduced $\chi^2$ is 1.51, compared to 1.65 for a fit to an absorbed power law without ALPs.
}
\label{signalFieldConfig}
\end{figure}

\begin{figure}
\centering
\includegraphics[width=0.6\textwidth]{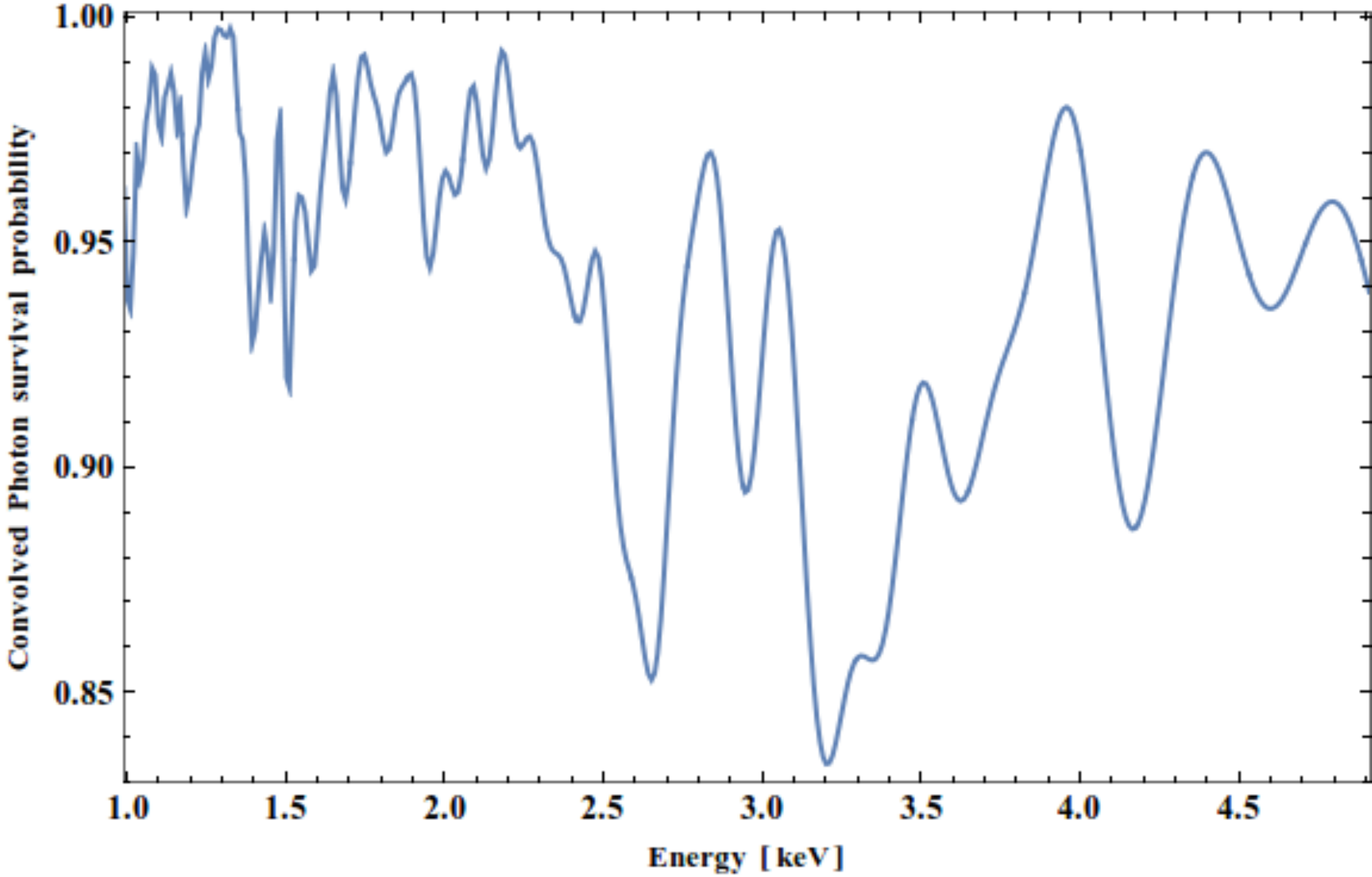}
\caption{The photon survival probability for the fit shown in Figure~\ref{signalFieldConfig}.
}
\label{signalFieldConfigProbs}
\end{figure}

It is not possible for us to constrain ALP couplings beyond a level where they would produce residuals comparable to these features.
We can estimate approximately that to produce such residuals would require an ALP-photon coupling of similar magnitude to the limits placed in the Section \ref{sec:bounds}: $g_{a \gamma \gamma} \sim 1 - 5 \times 10^{-12} \, {\rm GeV}^{-1}$. For example, Figure \ref{signalFieldConfig} shows a fit to the clean ACIS-I edge observations with an absorbed power law multiplied by the photon survival probability $P_{\gamma \to \gamma}$. In this case, $P_{\gamma \to \gamma}$ was calculated assuming the existence of ALPs with $g_{a \gamma \gamma} = 1 \times 10^{-12} \, {\rm GeV}^{-1}$ and a central magnetic field $B_0 = 25 \, \mu {\rm G}$ (the most optimistic field scenario used in Section \ref{sec:bounds}). We see that the anomalies at $2.2 \, {\rm keV}$ and $3.5 \, {\rm keV}$ have been alleviated by the presence of ALPs (although, for this magnetic
field, at the expense of creating similarly sized anomalies at higher energies). Figure \ref{signalFieldConfigProbs} shows the corresponding photon survival probability spectrum.

\section{Technical Details}

\subsection{The Presence of the 6.4 keV Iron K$\alpha$ line}
\label{sec:appb1}
As the cleaned spectra in the main body of the paper are only extracted up to 5 keV, it is a useful consistency check to ensure that
the reflected iron Fe K$\alpha$ line at 6.4 keV (6.3 keV after redshifting) is present in our spectra. For this
purpose we use the cleaned spectra for each
of the three sets of \emph{Chandra} observations, and fit the spectrum only between 5.5 and 7 keV.
We fit first a power-law, and then the sum of a
 power-law and a Gaussian (we do not fit $n_H$ as absorption is irrelevant at these energies).

\begin{figure}[h]
\begin{center}
\includegraphics[width=0.8\textwidth]{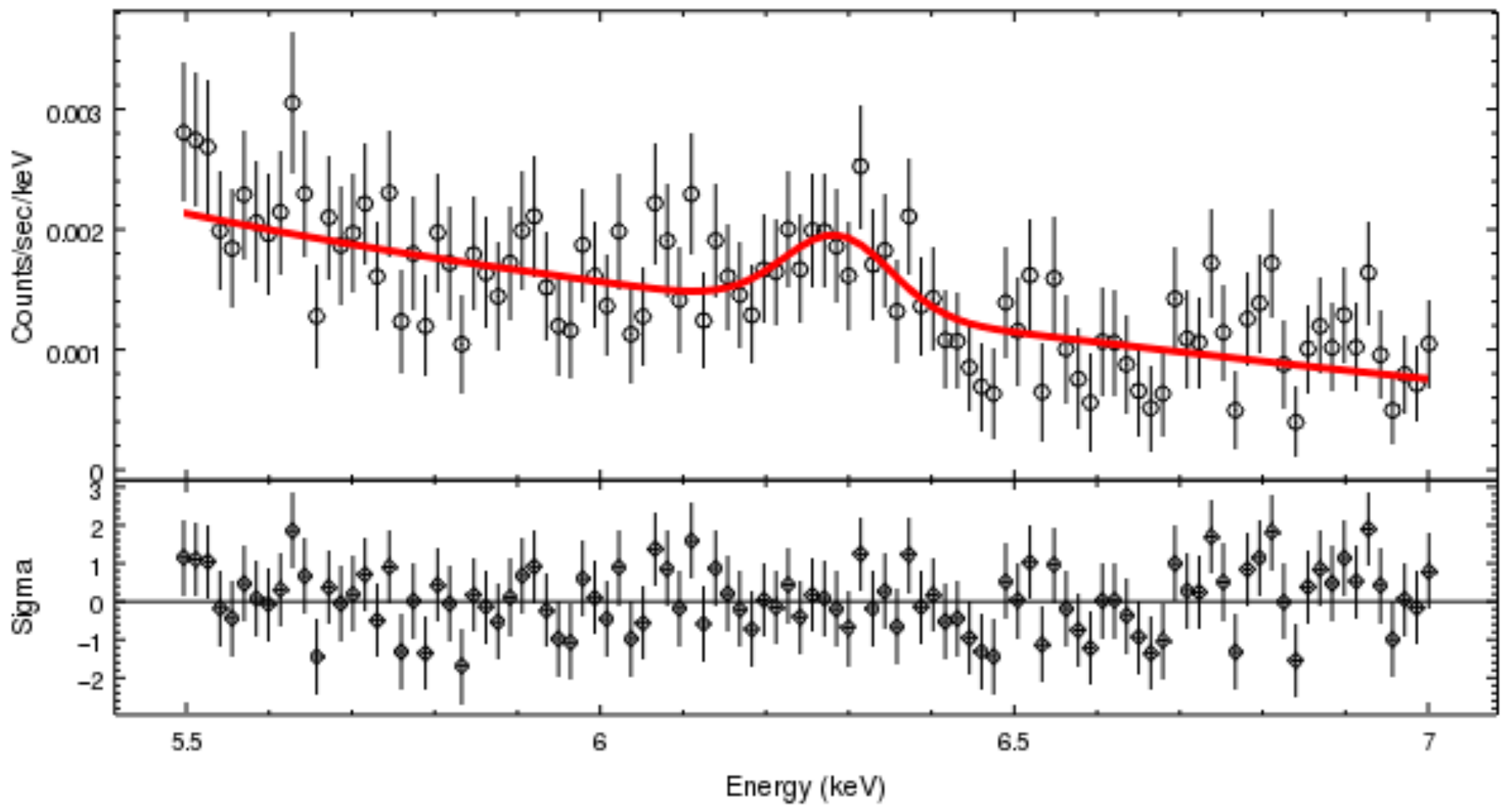}
\end{center}
\caption{The Fe K$\alpha$ line in the cleaned ACIS-S data.}\label{FeLine}
\end{figure}

 We determine the best-fit
 central energy of the additional Gaussian and the resulting improvement in $\chi^2$.
 These fits are summarised in Table~\ref{tab1} below and we also show explicitly the fit for the ACIS-S data in Figure~\ref{FeLine}.
  \begin{table}[h]
 \begin{center}
 \def\arraystretch{1.2}
 \begin{tabular}{c|c|c|c}

 Data set & Power-law index & $E_{\rm line}$ (keV) & $\Delta \chi^2$ (Degrees of freedom) \\
\hline \hline
ACIS-I Edge & $0.84_{-0.1}^{+0.3}$ & $6.32_{-0.05}^{+0.06}$ & 6.8/2 \\
\hline
ACIS-I Midway & $-0.14_{-0.02}^{+0.03}$ & $6.33_{-0.06}^{+0.05}$ & 2.2/2 \\
\hline
ACIS-S Central & $0.49_{-0.11}^{+0.07}$ & $6.29_{-0.03}^{+0.02}$ & 12.2/2
\end{tabular}
\caption{Fit parameters for a (power-law + Gaussian) fit between 5.5 and 7 keV.}\label{tab1}
\end{center}
\end{table}

These show that the iron line is clearly present in the ACIS-S and ACIS-I Edge data, and marginally present in the ACIS-I Midway data.
As ACIS-S has better energy resolution than ACIS-I, it is unsurprising that the Fe line is found at highest significance there.

We note that all spectral indices are unphysically
hard (and even negative for the ACIS-I Midway data). This is a clear consequence of pile-up. For energies above 5 keV, the \emph{Chandra} effective area
is falling off rapidly. There are also intrinsically far fewer source photons than are present at lower energies, given the power-law spectrum.
The effects of pile-up at these higher energies are therefore far more severe than at lower energies.

While these spectra have been cleaned, this does not mean that there is zero pile-up. For the bulk of the spectra, the
effects of pile-up in the cleaned spectra are at the level of a few per cent. However, for $E > 5~{\rm keV}$ the effects of pile-up are substantially more important for the above reasons: the combination of the intrinsic decline in the spectrum and the fall-off in the \emph{Chandra} effective area implies that pile-up plays a proportionally far
more severe role at high energies.

\subsection{Errors in the Fitted Parameters}
\label{sec:appb3}
For fits of spectra to models, the paper includes the \emph{fitted statistical errors}. However, significant caution should be applied when interpreting
such a quantity (for example, a power-law index) as an error on the intrinsic spectrum of the source. Systematic error on overall power-law indices are likely
to be much larger. While hard to quantify,
there are several clear sources of such systematic effects:
\begin{enumerate}

\item Although the cleaned spectra have significantly reduced levels of pile-up, no spectrum can be entirely free of pile-up. Pile-up automatically leads
to a hardening of spectral indices, as it moves photons from low to high energies.

\item
The point spread function of \emph{Chandra} is a function of energy, and more energetic photons tend to be spread out more. For spectra such as those used here,
where a central and highly piled-up core is excluded, this will bias the analysed photons to higher energy. While the \emph{Chandra} analysis software aims to take this into account,
this correction will not be perfect, particularly for off-axis sources where the form of the optical image is more complex (as for the ACIS-I Midway data, where the AGN image takes on a
`Maltese cross' form).

\item
Generally, there are systematic errors on the overall power-law index that will arise because the optical conditions are significantly different on-axis and off-axis. These are hard to quantify, but are certainly much larger than the purely statistical errors on fitted parameters.

\item
The AGN luminosity is time-variable and the power-law index will be time-variable as well. For observations taken at different times, the intrinsic power-law of the source
may be different.
\end{enumerate}

\bibliographystyle{aasjournal}
\bibliography{ngc1275bib}

\begin{thebibliography}{}
\expandafter\ifx\csname natexlab\endcsname\relax\def\natexlab#1{#1}\fi
\providecommand{\url}[1]{\href{#1}{#1}}

\bibitem[{Agrawal {et~al.}(2015)Agrawal, Chacko, Kilic, \&
  Verhaaren}]{1503.03057}
Agrawal, P., Chacko, Z., Kilic, C., \& Verhaaren, C.~B. 2015, JHEP, 08, 072

\bibitem[{Aharonian {et~al.}(2017)}]{Hitomi}
Aharonian, F.~A., {et~al.} 2017, Astrophys. J., 837, L15

\bibitem[{Alvarez {et~al.}(2015)Alvarez, Conlon, Day, Marsh, \&
  Rummel}]{14101867}
Alvarez, P.~D., Conlon, J.~P., Day, F.~V., Marsh, M. C.~D., \& Rummel, M. 2015,
  JCAP, 1504, 013

\bibitem[{Angus {et~al.}(2014)Angus, Conlon, Marsh, Powell, \&
  Witkowski}]{1312.3947}
Angus, S., Conlon, J.~P., Marsh, M. C.~D., Powell, A.~J., \& Witkowski, L.~T.
  2014, JCAP, 1409, 026

\bibitem[{Armengaud {et~al.}(2014)}]{1401.3233}
Armengaud, E., {et~al.} 2014, JINST, 9, T05002

\bibitem[{{Arnaud}(1996)}]{Arnaud}
{Arnaud}, K.~A. 1996, in Astronomical Society of the Pacific Conference Series,
  Vol. 101, Astronomical Data Analysis Software and Systems V, ed. G.~H.
  {Jacoby} \& J.~{Barnes}, 17

\bibitem[{Balmaverde {et~al.}(2006)Balmaverde, Capetti, \&
  Grandi}]{Balmaverde2006}
Balmaverde, B., Capetti, A., \& Grandi, P. 2006, Astron. Astrophys., 451, 35

\bibitem[{Berlin {et~al.}(2015)Berlin, DiFranzo, \& Hooper}]{1501.03496}
Berlin, A., DiFranzo, A., \& Hooper, D. 2015, Phys. Rev., D91, 075018

\bibitem[{Bonafede {et~al.}(2010)Bonafede, Feretti, Murgia, Govoni, Giovannini,
  Dallacasa, Dolag, \& Taylor}]{1002.0594}
Bonafede, A., Feretti, L., Murgia, M., {et~al.} 2010, Astron. Astrophys., 513,
  A30

\bibitem[{Boyarsky {et~al.}(2014)Boyarsky, Ruchayskiy, Iakubovskyi, \&
  Franse}]{Boyarsky}
Boyarsky, A., Ruchayskiy, O., Iakubovskyi, D., \& Franse, J. 2014, Phys. Rev.
  Lett., 113, 251301

\bibitem[{Brax {et~al.}(2015)Brax, Brun, \& Wouters}]{BraxWoutersBrun}
Brax, P., Brun, P., \& Wouters, D. 2015, Phys. Rev., D92, 083501

\bibitem[{Brockway {et~al.}(1996{\natexlab{a}})Brockway, Carlson, \&
  Raffelt}]{astro-ph/9605197}
Brockway, J.~W., Carlson, E.~D., \& Raffelt, G.~G. 1996{\natexlab{a}}, Phys.
  Lett., B383, 439

\bibitem[{Brockway {et~al.}(1996{\natexlab{b}})Brockway, Carlson, \&
  Raffelt}]{Brockway:1996yr}
---. 1996{\natexlab{b}}, Phys. Lett., B383, 439

\bibitem[{Bulbul {et~al.}(2014)Bulbul, Markevitch, Foster, Smith, Loewenstein,
  \& Randall}]{Bulbul}
Bulbul, E., Markevitch, M., Foster, A., {et~al.} 2014, Astrophys. J., 789, 13

\bibitem[{Burrage {et~al.}(2009)Burrage, Davis, \& Shaw}]{0902.2320}
Burrage, C., Davis, A.-C., \& Shaw, D.~J. 2009, Phys. Rev. Lett., 102, 201101

\bibitem[{{Carter} {et~al.}(2003){Carter}, {Karovska}, {Jerius}, {Glotfelty},
  \& {Beikman}}]{CHART}
{Carter}, C., {Karovska}, M., {Jerius}, D., {Glotfelty}, K., \& {Beikman}, S.
  2003, in Astronomical Society of the Pacific Conference Series, Vol. 295,
  Astronomical Data Analysis Software and Systems XII, ed. H.~E. {Payne}, R.~I.
  {Jedrzejewski}, \& R.~N. {Hook}, 477

\bibitem[{Churazov {et~al.}(2003)Churazov, Forman, Jones, \&
  Bohringer}]{Churazov:2003hr}
Churazov, E., Forman, W., Jones, C., \& Bohringer, H. 2003, Astrophys. J., 590,
  225

\bibitem[{Cicoli {et~al.}(2014)Cicoli, Conlon, Marsh, \& Rummel}]{14032370}
Cicoli, M., Conlon, J.~P., Marsh, M. C.~D., \& Rummel, M. 2014, Phys. Rev.,
  D90, 023540

\bibitem[{Cicoli {et~al.}(2012)Cicoli, Goodsell, \& Ringwald}]{1206.0819}
Cicoli, M., Goodsell, M., \& Ringwald, A. 2012, JHEP, 10, 146

\bibitem[{Cline \& Frey(2014{\natexlab{a}})}]{14080233}
Cline, J.~M., \& Frey, A.~R. 2014{\natexlab{a}}, JCAP, 1410, 013

\bibitem[{Cline \& Frey(2014{\natexlab{b}})}]{Cline}
---. 2014{\natexlab{b}}, Phys. Rev., D90, 123537

\bibitem[{Collaboration(2016)}]{1603.06978}
Collaboration, T. F.-L. 2016, arXiv:1603.06978

\bibitem[{Conlon(2006)}]{hep-th/0602233}
Conlon, J.~P. 2006, JHEP, 05, 078

\bibitem[{Conlon \& Day(2014)}]{14047741}
Conlon, J.~P., \& Day, F.~V. 2014, JCAP, 1411, 033

\bibitem[{Conlon \& Marsh(2013)}]{1305.3603}
Conlon, J.~P., \& Marsh, M. C.~D. 2013, Phys. Rev. Lett., 111, 151301

\bibitem[{Conlon {et~al.}(2015)Conlon, Marsh, \& Powell}]{1509.06748}
Conlon, J.~P., Marsh, M. C.~D., \& Powell, A.~J. 2015, arXiv:1509.06748

\bibitem[{Conlon \& Powell(2015)}]{14065518}
Conlon, J.~P., \& Powell, A.~J. 2015, JCAP, 1501, 019

\bibitem[{{Davis}(2001)}]{Davis:2001}
{Davis}, J.~E. 2001, Astrophys. J, 562, 575

\bibitem[{{Davis} {et~al.}(2012){Davis}, {Bautz}, {Dewey}, {Heilmann}, {Houck},
  {Huenemoerder}, {Marshall}, {Nowak}, {Schattenburg}, {Schulz}, \&
  {Smith}}]{MARX}
{Davis}, J.~E., {Bautz}, M.~W., {Dewey}, D., {et~al.} 2012, in \procspie, Vol.
  8443, Space Telescopes and Instrumentation 2012: Ultraviolet to Gamma Ray,
  84431A

\bibitem[{D'Eramo {et~al.}(2016)D'Eramo, Hambleton, Profumo, \&
  Stefaniak}]{Profumo}
D'Eramo, F., Hambleton, K., Profumo, S., \& Stefaniak, T. 2016,
  arXiv:1603.04859

\bibitem[{Dobrynina {et~al.}(2015)Dobrynina, Kartavtsev, \&
  Raffelt}]{1412.4777}
Dobrynina, A., Kartavtsev, A., \& Raffelt, G. 2015, Phys. Rev., D91, 083003

\bibitem[{Fabian {et~al.}(2015)Fabian, Walker, Pinto, Russell, \&
  Edge}]{Fabian:2015kua}
Fabian, A.~C., Walker, S.~A., Pinto, C., Russell, H.~R., \& Edge, A.~C. 2015,
  Mon. Not. Roy. Astron. Soc., 451, 3061

\bibitem[{Fairbairn {et~al.}(2011)Fairbairn, Rashba, \& Troitsky}]{0901.4085}
Fairbairn, M., Rashba, T., \& Troitsky, S.~V. 2011, Phys. Rev., D84, 125019

\bibitem[{Finkbeiner \& Weiner(2014)}]{Finkbeiner}
Finkbeiner, D.~P., \& Weiner, N. 2014, arXiv:1402.6671

\bibitem[{{Freeman} {et~al.}(2001){Freeman}, {Doe}, \&
  {Siemiginowska}}]{sherpa}
{Freeman}, P., {Doe}, S., \& {Siemiginowska}, A. 2001, in Proceedings of the
  International Society for Optical Engineering, Vol. 4477, Astronomical Data
  Analysis, ed. J.-L. {Starck} \& F.~D. {Murtagh}, 76--87

\bibitem[{{Fruscione} {et~al.}(2006){Fruscione}, {McDowell}, {Allen},
  {Brickhouse}, {Burke}, {Davis}, {Durham}, {Elvis}, {Galle}, {Harris},
  {Huenemoerder}, {Houck}, {Ishibashi}, {Karovska}, {Nicastro}, {Noble},
  {Nowak}, {Primini}, {Siemiginowska}, {Smith}, \& {Wise}}]{ciao}
{Fruscione}, A., {McDowell}, J.~C., {Allen}, G.~E., {et~al.} 2006, in
  Proceedings of the International Society for Optical Engineering, Vol. 6270,
  Society of Photo-Optical Instrumentation Engineers (SPIE) Conference Series,
  62701V

\bibitem[{Grifols {et~al.}(1996{\natexlab{a}})Grifols, Masso, \&
  Toldra}]{astro-ph/9606028}
Grifols, J.~A., Masso, E., \& Toldra, R. 1996{\natexlab{a}}, Phys. Rev. Lett.,
  77, 2372

\bibitem[{Grifols {et~al.}(1996{\natexlab{b}})Grifols, Masso, \&
  Toldra}]{Grifols:1996id}
---. 1996{\natexlab{b}}, Phys. Rev. Lett., 77, 2372

\bibitem[{Harris {et~al.}(2006)Harris, Cheung, Biretta, Sparks, Junor, Perlman,
  \& Wilson}]{0004-637X-640-1-211}
Harris, D.~E., Cheung, C.~C., Biretta, J.~A., {et~al.} 2006, The Astrophysical
  Journal, 640, 211.
\newblock \url{http://stacks.iop.org/0004-637X/640/i=1/a=211}

\bibitem[{Horns {et~al.}(2012)Horns, Maccione, Meyer, Mirizzi, Montanino, \&
  Roncadelli}]{1207.0776}
Horns, D., Maccione, L., Meyer, M., {et~al.} 2012, Phys. Rev., D86, 075024

\bibitem[{Iakubovskyi(2015)}]{1510.00358}
Iakubovskyi, D. 2015, arXiv:1510.00358

\bibitem[{Meyer {et~al.}(2013)Meyer, Horns, \& Raue}]{1302.1208}
Meyer, M., Horns, D., \& Raue, M. 2013, Phys. Rev., D87, 035027

\bibitem[{Mirizzi \& Montanino(2009)}]{0911.0015}
Mirizzi, A., \& Montanino, D. 2009, JCAP, 0912, 004

\bibitem[{Morrison \& McCammon(1983)}]{Morrison:1983hg}
Morrison, R., \& McCammon, D. 1983, Astrophys. J., 270, 119

\bibitem[{Payez {et~al.}(2012)Payez, Cudell, \& Hutsemekers}]{1204.6187}
Payez, A., Cudell, J.~R., \& Hutsemekers, D. 2012, JCAP, 1207, 041

\bibitem[{Payez {et~al.}(2015)Payez, Evoli, Fischer, Giannotti, Mirizzi, \&
  Ringwald}]{1410.3747}
Payez, A., Evoli, C., Fischer, T., {et~al.} 2015, JCAP, 1502, 006

\bibitem[{Powell(2015)}]{1411.4172}
Powell, A.~J. 2015, JCAP, 1509, 017

\bibitem[{Raffelt \& Stodolsky(1988)}]{Raffelt:1987im}
Raffelt, G., \& Stodolsky, L. 1988, Phys. Rev., D37, 1237

\bibitem[{Ringwald(2012)}]{RingwaldReview}
Ringwald, A. 2012, Phys. Dark Univ., 1, 116

\bibitem[{Schlederer \& Sigl(2016)}]{Schlederer}
Schlederer, M., \& Sigl, G. 2016, JCAP, 1601, 038

\bibitem[{Sikivie(1983)}]{Sikivie:1983ip}
Sikivie, P. 1983, Phys. Rev. Lett., 51, 1415, [Erratum: Phys. Rev.
  Lett.52,695(1984)]

\bibitem[{Sikivie(1985)}]{Sikivie:1985}
---. 1985, Phys. Rev. D, 32, 2988.
\newblock \url{http://link.aps.org/doi/10.1103/PhysRevD.32.2988}

\bibitem[{Svrcek \& Witten(2006)}]{hep-th/0605206}
Svrcek, P., \& Witten, E. 2006, JHEP, 06, 051

\bibitem[{Tavecchio {et~al.}(2012)Tavecchio, Roncadelli, Galanti, \&
  Bonnoli}]{1202.6529}
Tavecchio, F., Roncadelli, M., Galanti, G., \& Bonnoli, G. 2012, Phys. Rev.,
  D86, 085036

\bibitem[{Taylor {et~al.}(2006)Taylor, Gugliucci, Fabian, Sanders, Gentile, \&
  Allen}]{0602622}
Taylor, G.~B., Gugliucci, N.~E., Fabian, A.~C., {et~al.} 2006, Mon. Not. Roy.
  Astron. Soc., 368, 1500

\bibitem[{Vacca {et~al.}(2012)Vacca, Murgia, Govoni, Feretti, Giovannini,
  Perley, \& Taylor}]{Vacca:2012up}
Vacca, V., Murgia, M., Govoni, F., {et~al.} 2012, Astron. Astrophys., 540, A38

\bibitem[{Wouters \& Brun(2012)}]{1205.6428}
Wouters, D., \& Brun, P. 2012, Phys. Rev., D86, 043005

\bibitem[{Wouters \& Brun(2013)}]{1304.0989}
---. 2013, Astrophys. J., 772, 44

\bibitem[{{Yamazaki} {et~al.}(2013){Yamazaki}, {Fukazawa}, {Sasada}, {Itoh},
  {Nishino}, {Takahashi}, {Takaki}, {Kawabata}, {Yoshida}, \&
  {Uemura}}]{Yamazaki}
{Yamazaki}, S., {Fukazawa}, Y., {Sasada}, M., {et~al.} 2013, Publications of
  the Astronomical Society of Japan, 65, doi:10.1093/pasj/65.2.30

\end{thebibliography}

\end{document}